\documentclass[a4paper,12pt]{article} 
\input{epsf.tex}
\usepackage{amssymb,epsfig}

\newcommand{\One}{1\kern-4.5pt1}

\newcommand {\beq}{\begin{equation}}
\newcommand {\eeq}{\end{equation}}
\newcommand {\bea}{\begin{eqnarray}}
\newcommand {\eea}{\end{eqnarray}}

\begin{document}

\addtolength{\baselineskip}{0.20\baselineskip}

\hfill August 2008

\vspace{48pt}

\centerline{\Huge Non-compact QED$_3$ at finite temperature:}
\centerline{\Huge the confinement-deconfinement transition}
\vspace{18pt}

\centerline{\bf Roberto Fiore$^a$, Pietro Giudice$^b$ and Alessandro Papa$^a$} 

\vspace{15pt}

\centerline{$^a$ \sl Dipartimento di Fisica, Universit\`a della Calabria}
\centerline{\sl and INFN, Gruppo collegato di Cosenza,}
\centerline{\sl I-87036 Arcavacata di Rende, Cosenza, Italy.}

\centerline{$^b$ \sl Dipartimento di Fisica Teorica, Universit\`a di Torino} 
\centerline{\sl and INFN, Sezione di Torino,} 
\centerline{\sl via P. Giuria 1, I-10125 Torino, Italy.}
\smallskip

\vspace{24pt}

\centerline{{\bf Abstract}}

\noindent
{\narrower
The confinement-deconfinement phase transition is explored by lattice 
numerical simulations in non-compact (2+1)-dimensional quantum electrodynamics 
with massive fermions at finite temperature. The existence of two phases, one
with and the other without confinement of fractional charges, is related to 
the realization of the $\mathbb{Z}$ symmetry. The order parameter of this 
transition can be clearly identified.
We show that it is possible to detect the critical temperature for a 
given value of the fermion mass, by exploiting suitable lattice
operators as probes of the $\mathbb{Z}$ symmetry.
Moreover, the large-distance behavior of the correlation of these
operators permits to distinguish the phase with Coulomb-confinement from 
the Debye-screened phase. The resulting scenario is compatible
with the existence of a Berezinsky-Kosterlitz-Thouless transition.
Some investigations are presented on the possible relation between chiral 
and deconfinement transitions and on the role of ``monopoles''.
}



\vfill

\newpage

\section{Introduction} 

Non-compact quantum electrodynamics in (2+1)-dimensions (QED$_3$) with
fermionic matter is a special theory since it plays a role in contexts 
ranging from condensed matter to particle physics. In particular, this theory 
is relevant for the characterization of the phase diagram of high-$T_c$ 
superconductors in the temperature-doping plane 
(see, for instance, Ref.~\cite{Mavromatos:2003ss} and references therein). 
Moreover, it is an interesting theoretical laboratory for the
investigation of mechanisms of confinement and for the study of the
confinement-deconfinement transition at finite temperature.

This paper aims at contributing to clarify some issues related with the latter
topic. Understanding the mechanism of confinement, identifying the nature 
of the confinement-deconfinement transition and the related order parameter 
is of central interest in finite-temperature non-Abelian quantum field 
theories, such as Quantum Chromodynamics (QCD). Here definite answers are
available in the limiting cases of infinite quark masses (i.e. in the pure
gauge theory) and zero quark masses (i.e. in the chiral limit). Instead, many 
open questions remain for the general case of fermions with finite non-zero 
mass. In non-compact QED$_3$ the symmetry whose breaking determines the
transition is the $\mathbb{Z}$ symmetry, independently from the fermionic
mass, the order parameter being the Polyakov loop with fractional charge.
Evaluating the vacuum expectation value of the order parameter and
its 2-point correlator allows to get, at least in principle, a complete 
description of the phase diagram of the theory for varying fermion masses,
coupling constant and temperature. 

The only theoretical scenario suggested so far for the phase diagram
of non-compact QED$_3$ at finite temperature is presented in  
Refs.~\cite{Grignani:1995iv,Grignani:1995hx},
where it is found that the effective action for the temporal component of 
the gauge field, $A_0(x)$, in the limit of large fermion mass becomes 
equivalent to the sine-Gordon potential. This led the authors of 
Ref.~\cite{Grignani:1995iv} to conclude that there is a phase transition 
of Berezinsky-Kosterlitz-Thouless (BKT) 
type~\cite{Berezinsky:1970fr,Kosterlitz:1973xp}. More precisely, there is 
a critical temperature $T_c$, depending on the fermion mass, above which 
the system is in a deconfined or Debye phase, while below $T_c$ the 
interaction is logarithmic with the distance, i.e. it is Coulomb-like.
In this paper we intend to verify the above scenario for non-compact QED$_3$ 
at finite temperature, by Monte Carlo numerical simulations, using as
probes the Polyakov loop with fractional charge and its 2-point correlator.

Another topic we consider in this paper is the behavior of the chiral
condensate across the deconfinement transition. In particular, if the 
chiral condensate exhibits a sharp drop when the temperature is increased 
through the critical value, then we can conclude that the dynamical 
mechanism responsible for deconfinement affects also the chiral symmetry
of the theory.

Finally, we present a few numerical results concerning the magnetic monopole
density. While monopoles in compact QED$_3$ are undoubtedly related to 
confinement in the pure gauge case~\cite{polyakov} and the same is argued
also in presence of fermions~\cite{Herbut:2003bs,Azcoiti:1993ey,Fiore:2005ps},
in the non-compact theory there is no {\it a priori} reason for which they
could play any role. However, following Ref.~\cite{Hands:1989cg}, a monopole 
density can be defined on the lattice in the non-compact theory
exactly as in the compact one and its behavior with temperature can
be studied. In (3+1)-dimensional non-compact QED, it turns out that there
exist a percolation threshold for monopole current networks near the chiral 
transition~\cite{Hands:1989cg}, thus suggesting a possible relation between
monopole condensation and chiral symmetry breaking (see 
Ref.~\cite{Rakow:1992ae} for a criticism to this approach). In this paper we 
present some determinations for the monopole density across the deconfinement 
transition and briefly discuss the possibility of a relation between monopoles
and confinement in non-compact QED$_3$.

The paper is organized as follows: in Section~\ref{sec:continuum} we briefly
recall the theory in the continuum formulation, the origin of the $\mathbb{Z}$ 
symmetry and the conjectured phase diagram; in Section~\ref{sec:lattice} 
we describe the lattice version of the theory and the operators we use as 
probes of the phase transition; in Section~\ref{sec:results} we present
the numerical results and discuss their interpretation; in 
Section~\ref{sec:conclusions} we draw our conclusions and sketch the future
perspectives. 

\section{The continuum QED$_3$ theory}
\label{sec:continuum}

The continuum Lagrangian density describing QED$_3$ is given in
Minkowski metric by
\begin{equation}
{\cal L} =-\frac{1}{4} F_{\mu \nu}^2+\overline\psi_i iD_\mu \gamma^\mu \psi_i -
 m_0\overline\psi_i \psi_i \;,
\end{equation}
where $F_{\mu \nu}=\partial_\mu A_\nu - \partial_\nu A_\mu$ is the field 
strength, $D_\mu=\partial_\mu-igA_\mu$ is the covariant derivative, $g$ is 
the coupling constant (or the electric charge) and the fermion fields 
$\psi_i$ ($i=1, \dots, N_f$) are 4-component spinors, defined 
in such a way that the theory is parity-invariant\footnote{This implies
that the fermion is massive, while the photon is massless.}. 
The relevant information concerning parity and chiral symmetry of this model 
can be found, for instance, in Ref.~\cite{Fiore:2005ps}, Section II.

\subsection{The $\mathbb{Z}$ symmetry}
\label{subsec:Z}

In this Section we briefly recall the origin of the $\mathbb{Z}$ symmetry
which plays a fundamental role in this paper.

The Euclidean partition function of finite temperature QED is invariant
under gauge transformations~\cite{Grignani:1995iv,Grignani:1995hx}.
For the field $A_\mu(\tau,\vec x)$ the gauge transformation 
\beq
A^\prime_\mu(\tau,\vec{x})=A_\mu(\tau,\vec{x})+\partial_\mu\chi(\tau,\vec{x}) \;,
\label{eq_gt_A}
\eeq
together with the periodic boundary conditions in the time direction 
\beq
A_\mu(1/T,\vec{x})=A_\mu(0,\vec{x})\;,
\eeq
where $T$ is the temperature, implies the following condition:
\beq
\partial_\mu\chi(1/T,\vec{x})=\partial_\mu\chi(0,\vec{x})\;.
\eeq
Alike, for the fermion field $\psi(\tau,\vec{x})$, the gauge transformation
\beq
\psi^\prime(\tau,\vec{x})=e^{ig\chi(\tau,\vec{x})}\psi(\tau,\vec{x})\;,
\eeq
together with antiperiodic boundary conditions in the time direction 
\beq
\psi(1/T,\vec{x})=-\psi(0,\vec{x})\;,
\eeq
implies
\beq
\chi(1/T,\vec{x})=\chi(0,\vec{x})+\frac{2\pi}{g}n \;,
\label{eq_gft}
\eeq
where $n$ is an integer. Differently, we can say that $\chi(x)$ is 
periodic in the time direction with period $1/T$ up to an integer multiple
of $2\pi/g$. In the language of group theory, we can say that if $G$ is the 
group of all gauge transformations and $H$ is the
subgroup of those gauge transformations which are strictly periodic, then
the quotient group $G/H$ is isomorphic to $\mathbb{Z}$, the additive 
group of integers. We refer to this when we say that the theory possesses 
$\mathbb{Z}$ symmetry. Note that this is a symmetry of the partition function 
\emph{in presence of dynamical electrons}.

In order to study the realization of the $\mathbb{Z}$ symmetry, a version 
of the Polyakov loop operator is introduced, whose average is related to 
the free energy of an external charge $\tilde{g}$:
\beq
\Pi_{\tilde g}(\vec{x}) = e^{i \tilde{g} \int_0^{1/T} dx_0 A_0(x_0,\vec{x})}\;.
\label{eq-pl}
\eeq
Under the action of an element of $G/H$,
the Polyakov loop operator with charge $\tilde{g}$ transforms as (here we use 
Eqs.(\ref{eq_gt_A}) and~(\ref{eq_gft}))
\beq
\Pi^\prime_{\tilde g}(\vec{x}) =e^{i \tilde{g} \int_0^{1/T} d\tau 
A_0^\prime(\tau,\vec{x})}=
e^{i \tilde{g} \int_0^{1/T} d\tau A_0(\tau,\vec{x})} e^{\frac{i 2 \pi n 
\tilde{g}}{g}}=
\Pi_{\tilde g}(\vec{x}) e^{\frac{i 2 \pi n \tilde{g}}{g}}\;.
\label{eq-pl-0}
\eeq
Therefore, in order that the operator defined in Eq.~(\ref{eq-pl}) be an 
order parameter, the charge $\tilde g$ must not be an integer multiple of 
the basic charge $g$. One possibility is to follow Ref.~\cite{Hands:2002dv} 
and study charges that are rational fractions of the fundamental charge, 
i.e. $\tilde{g}=g/m$, with integer $m$. As a consequence, Eq.~(\ref{eq-pl-0})
becomes
\beq
\Pi'_{m}(\vec{x}) = \Pi_{m}(\vec{x}) e^{\frac{i 2 \pi n}{m}}\;.
\eeq

The realization of the $\mathbb{Z}$ symmetry tests the ability of the 
electrodynamic system to screen charges which are not integral multiples
of the electron charge. Therefore, it is related to (fractional) charge 
screening and confinement in QED in the same way as the $\mathbb{Z}_N$ 
symmetry does in finite temperature $SU(N)$ pure gauge theory.

In conclusion, the operator $\Pi_{\tilde g}(\vec{x})$ can be used as 
an order parameter for confinement in Abelian gauge theories even in 
presence of dynamical charged particles: if the symmetry is unbroken, the 
loop operator averages to zero and the system is in the confining phase, 
otherwise it is in a non-confining phase.

\subsection{Theoretical expectation for the phase diagram}
\label{subsec:phase}

In this Subsection, we briefly recall the ideas put forward in 
Ref.~\cite{Grignani:1995iv} about the phase structure of the theory
under consideration.

The authors of Ref.~\cite{Grignani:1995iv} propose the Polyakov 
loop with fractional charge as order parameter and argue that in
parity-invariant electrodynamics with fermions of mass $M$ there is a 
deconfinement transition at finite temperature of BKT type.

The analysis of Ref.~\cite{Grignani:1995iv} is based on computing
the effective action $V(M,gA_0/T)$ for $A_0(\tau,\vec{x})$, which 
explicitly exhibits the global $\mathbb{Z}$ symmetry. By studying 
this effective action they characterize the type of the phase transition.

At finite temperature, non-compact QED$_3$ contains three parameters
(with the dimension of mass): the fermion mass $M$, the gauge coupling 
$g^2$ and the temperature $T$; the dimensionless parameter which governs 
the loop expansion is the smaller between $g^2/M$ and $g^2/T$.

In the large $M$ limit, so that $T/M$ and $g^2/M$ are small with $g^2/T$ 
finite, they argue that the critical behavior of the theory is identical to 
that of the 2-dimensional sine-Gordon potential, which undergoes a BKT phase 
transition. Therefore they conclude that also in QED$_3$ at finite temperature 
there must be a BKT transition with a critical line in the $[M/T,g^2/T]$ 
plane, starting at
\beq
(M/T,g^2/T)=(\infty,8\pi)\;,
\eeq
from which they obtain
\beq
T_{crit.}^{M \gg T}=\frac{g^2}{8\pi}\;.
\label{t_crit}
\eeq
They determine also the critical temperature for the BKT transition up
to one-loop order:
\beq
T_{crit.}^{M \gg T}=\frac{g^2}{8 \pi (1+\frac{g^2}{12\pi M})}\;.
\eeq

The second limit considered in Ref.~\cite{Grignani:1995iv} is the 
high-temperature limit, $T \gg M,g^2$. The analysis of 
Ref.~\cite{Grignani:1995iv} shows that the $\mathbb{Z}$ symmetry is 
spontaneously broken, which means the system is in the deconfined phase.

In a subsequent paper~\cite{Grignani:1995hx} the same authors 
characterize better the nature of the two phases divided by the BKT critical 
line. They find that, where the $\mathbb{Z}$ symmetry is unbroken, the system 
is in a confining phase, more precisely in a ``Coulomb phase'', where 
electric charges mutually interact by a logarithmic Coulomb potential. 
This logarithmic behavior has as a consequence a power law dependence of the 
Polyakov loop correlators. Indeed, if the interaction potential is of the
form $V(r)=\alpha \log{r}$, then one can write immediately 
\beq
G(r)=\langle \Pi(0)\Pi^*(r) \rangle  =e^{-\frac{V(r)}{T}}=e^{-\frac{\alpha}{T} 
\log{r}}=r^{-\frac{\alpha}{T}}=r^{-\eta(T)}\;.
\label{correl_0}
\eeq
At the tree level, they find
\beq
G(r)_{\mbox{\scriptsize tree}} \propto r^{-\frac{\tilde{g}^2}{4 \pi T}}\;.
\label{correl_1}
\eeq
From Eqs.~(\ref{correl_0}) and~(\ref{correl_1}) the value of $\eta$ 
can be easily found:
\beq
\eta=\frac{\tilde{g}^2}{4 \pi T} \;.
\label{eta}
\eeq

Above $T_c$ the electric charge is not confined and the system is in a 
Debye plasma phase. The expected {\em large distance} behavior for the
{\em connected} Polyakov loop correlators in this case is given by
\beq
G_{\mbox{\scriptsize conn}}(r) \propto e^{-M_Dr} \;.
\eeq
The Debye mass $M_D$ makes the Coulomb interaction short-ranged and fermions 
and antifermions are approximately free particles.

\section{The lattice QED$_3$ theory}
\label{sec:lattice}

In this paper we discretize the Euclidean action on a lattice 
with $N_\sigma^2\times N_\tau$ sites and lattice spacing $a$ by staggered 
fermions $\overline\chi,\chi$, according to
\begin{equation}
\label{action}
 S=S_G+\sum_{i=1}^{N} \sum_{n,m}  \overline\chi_i(n)
M_{n,m} \chi_i(m)\;,
\end{equation}
where $S_G$ is the gauge field action and the fermion matrix is given by
\begin{equation}
M_{n,m}[U]=\sum_{\nu=1,2,3} \frac{\eta_{\nu}(n)}{2} \left\{ [U_{\nu}(n)]  
\delta_{m,n+\hat{\nu}}-[U_{\nu}^{\dagger}(m)] \delta_{m,n-\hat{\nu}} \right\}
\;,
\end{equation}
with $\eta_\nu(n)=(-1)^{\sum_{\rho<\nu} n_\rho}$ and 
$U_{\mu}(n)=e^{i \alpha_{\mu}(n)}$ is the link variable; the phase $\alpha$
is related to the gauge field by $\alpha_{\mu}(n)=a g A_{\mu}(n)$.

In a non-compact formulation of QED$_3$, $S_G$ is given by
\begin{equation}
S_G[\alpha]= \frac{\beta}{2} \sum_{n,\mu < \nu} F_{\mu \nu}(n)F_{\mu \nu}(n)\;,
\end{equation}
where
\begin{equation}
F_{\mu \nu}(n)=  \{ \alpha_{\nu}(n+\hat{\mu})-
  \alpha_{\nu}(n) \} - \{ \alpha_{\mu}(n+\hat{\nu})-
  \alpha_{\mu}(n) \}
\end{equation}
and $\alpha_{\mu}(n)$ is the phase of the link variable and $\beta=1/(g^2 a)$.

With this action we simulate $N=1$ flavors of staggered fermions corresponding
to $N_f=2$ flavors of parity-invariant four-component fermions.

\subsection{Order parameters and other lattice operators} 
\label{subsec:operators}

The lattice version of the operator~(\ref{eq-pl}) is~\cite{Hands:2002dv}
\begin{equation}
\Pi_m(\vec x)=\prod_{\tau=1}^{N_{\tau}} e^{\frac{i}{m} 
\theta(\stackrel{\rightarrow}{x},\tau)}\;,
\end{equation}
where $\theta=g A_0$ and $\vec x$ lives on the $N_\sigma^2$ spatial lattice,
from which we can determine the operator averaged on the lattice configuration,
\begin{equation}
\Pi_m  =  \frac{1}{N_\sigma^2} \sum_{\vec x}\Pi_m(\vec x)\;.
\label{pi_m}
\end{equation}
As noted in Ref.~\cite{Hands:2002dv}, the original $\mathbb{Z}$ symmetry is 
translated into a $\mathbb{Z}_m$ symmetry on the lattice operator $\Pi_m$; 
in the broken phase, the operator $\Pi_m$ will fluctuate around 
the values $e^{i \frac{2\pi}{m}k }$, where $k=0,1, \dots, m-1$.
In order to study the breaking of this symmetry it is useful therefore
to introduce the following operator: 
\begin{equation}
\Pi_m^m  =\left(  \frac{1}{N_\sigma^2} \sum_{\vec x}
\Pi_m(\vec x) \right)^m\;.
\label{pi_m_m}
\end{equation}
Another possible choice of lattice order parameter is~\cite{Gottlieb:1985ug}
\beq
\Theta = \cos \left[ m \times \arg \left( \Pi_m \right) \right]\;;
\label{cos_m}
\eeq
both definitions (\ref{pi_m_m}) and (\ref{cos_m}) have the effect to
``rotate'' the $m$ different phases to the direction corresponding
to $\theta=0$.

There are two other quantities of physical relevance to be considered; one
is the monopole density $\rho$, which can be defined in the same way as in 
the compact theory, using the method of Ref.~\cite{DeGrand:1980eq},
\begin{equation}
\rho=\frac{1}{2} \frac{\langle N_M \rangle + \langle N_{\bar{M}} 
\rangle}{N_\sigma^2 N_\tau}\;,
\end{equation}
where $N_M$ ($N_{\bar{M}}$) is the number of monopoles (antimonopoles);
the other is chiral condensate, $\langle \bar\chi \chi \rangle$. 

The presence of transitions is detected by looking for peaks in the 
susceptibility of the operators (\ref{pi_m}),~(\ref{pi_m_m}) and 
(\ref{cos_m}), the susceptibility of a generic operator $\cal O$ being 
defined as
\begin{equation}
\chi_{\cal O}=\langle {\cal O}^2 \rangle - \langle {\cal O} \rangle^2\;.
\end{equation}
The susceptibilities of monopole density and chiral condensate
is also considered in order to study the possible relation
of these latter operators with the confinement/deconfinement transition.

\subsection{General strategy of the lattice calculation}
\label{subsec:strategy}

The main goal of the numerical computations in this paper is to find
evidences of the existence of a transition between two phases, one with
unbroken and the other with broken $\mathbb{Z}$ symmetry. The strategy for 
that is to scan the temperature for a fixed value of the bare fermion mass 
$aM$ and on lattices with given extension to study the behavior of the lattice
order parameters defined in the previous subsection and of their 
susceptibility. On the lattice, we have
\begin{equation}
T=\frac{1}{N_\tau a}=\frac{g^2}{N_\tau}\beta\;,
\label{lattice_T}
\end{equation}
therefore at fixed coupling constant $g$ and on a lattice with fixed $N_\tau$, 
the temperature can be changed by changing $\beta$ (note that the theory
is super-renormalizable, therefore $g$ does not depend on $a$).
We expect that changing $\beta$ we meet a clear signature of the transition, 
in correspondence to the critical temperature, which, at infinite fermion
mass, is given in Eq.~(\ref{t_crit}) and, through~(\ref{lattice_T}),
can be translated on the lattice to the following critical value for the 
$\beta$ parameter:
\beq
\beta_c(M\to\infty)=\frac{N_{\tau}}{8\pi}\;.
\label{betac}
\eeq

The next step is to study the behavior of the Polyakov loop correlator.
If the theoretical expectations introduced in Section~\ref{subsec:phase} are 
true, then for a $\beta$ value below the critical one, $\beta_c(M)$, we 
should find a power law behavior, whereas above $\beta_c(M)$ there should be 
exponential fall-off. In the Coulomb-confined phase, i.e. for 
$\beta<\beta_c(M)$, where the power law behavior is expected, we can compare 
our findings with the expected tree-level value of $\eta$ given in 
Eq.~(\ref{eta}), which in lattice units reads
\beq
\eta=\frac{N_{\tau}}{m^2 4 \pi \beta} \;.
\label{etalattice}
\eeq
We recall that $1/m$ represents the ratio $\tilde g/g$, i.e. the
fraction charge in units of the fundamental charge. It should be noted 
here that $\eta$ is proportional to $1/m^2$.

\section{Numerical results}
\label{sec:results}

In this Section we present our numerical results obtained on lattices 
$12^2 \times 8$, $32^2 \times 8$ and $64^2 \times 8$. On the smallest
lattice considered, we found that tunneling effects among the different 
$\mathbb{Z}_m$ vacua of the theory in the broken symmetry phase are evident 
in the expectation value of $\Pi_m$. For this observable this makes possible
to better unravel the vacuum structure in the deconfined phase.
However, for the study of correlation functions, where tunneling in the
deconfined phase is potentially dangerous, we used lattices with 
$N_\sigma=32$ and 64 for which we verified that tunneling is absent.

\subsection{The algorithm}
\label{subsec:algorithm}

We used the Hybrid Monte Carlo Algorithm~\cite{Duane:1987de}, 
built by superimposing the Metropolis 
acceptance test to the Hybrid Algorithm in the structure proposed in 
Ref.~\cite{Duane:1986iw}). Here is a list of the internal parameters of 
the Monte Carlo code:
\begin{itemize}
\item
the mass term $\omega$ for the pseudofermionic field in the Hamiltonian, which 
drives the molecular dynamics; 
\item the integration step $\delta t$, fixed to 0.03 in all simulations;
\item the frequency MCR of the algorithm refreshments and Metropolis tests.
\end{itemize}
We have verified the important role of $\omega$ in simulations:
if $\omega$ is too small ($\omega\ll 1$), the system thermalizes very slowly,
but observables do not fluctuate much around the mean value during the 
simulation time; on the contrary, if $\omega$ is not so small 
($\omega \lesssim 1$), the system thermalizes very quickly with the 
disadvantage that fluctuations increase considerably. It is therefore
fundamental to optimize the choice of the parameter $\omega$.

Additionally, we use the following abbreviations: FOM for the frequency 
of measurements and NMS for the number of measurements or statistics. 
The bare fermion mass is fixed to $aM=0.05$.

\subsection{Order parameter}
\label{subsec:parameter}

\begin{figure}[htbp]
\centering
\hspace{-1.3cm}
\includegraphics[width=0.75\textwidth,angle=0]
{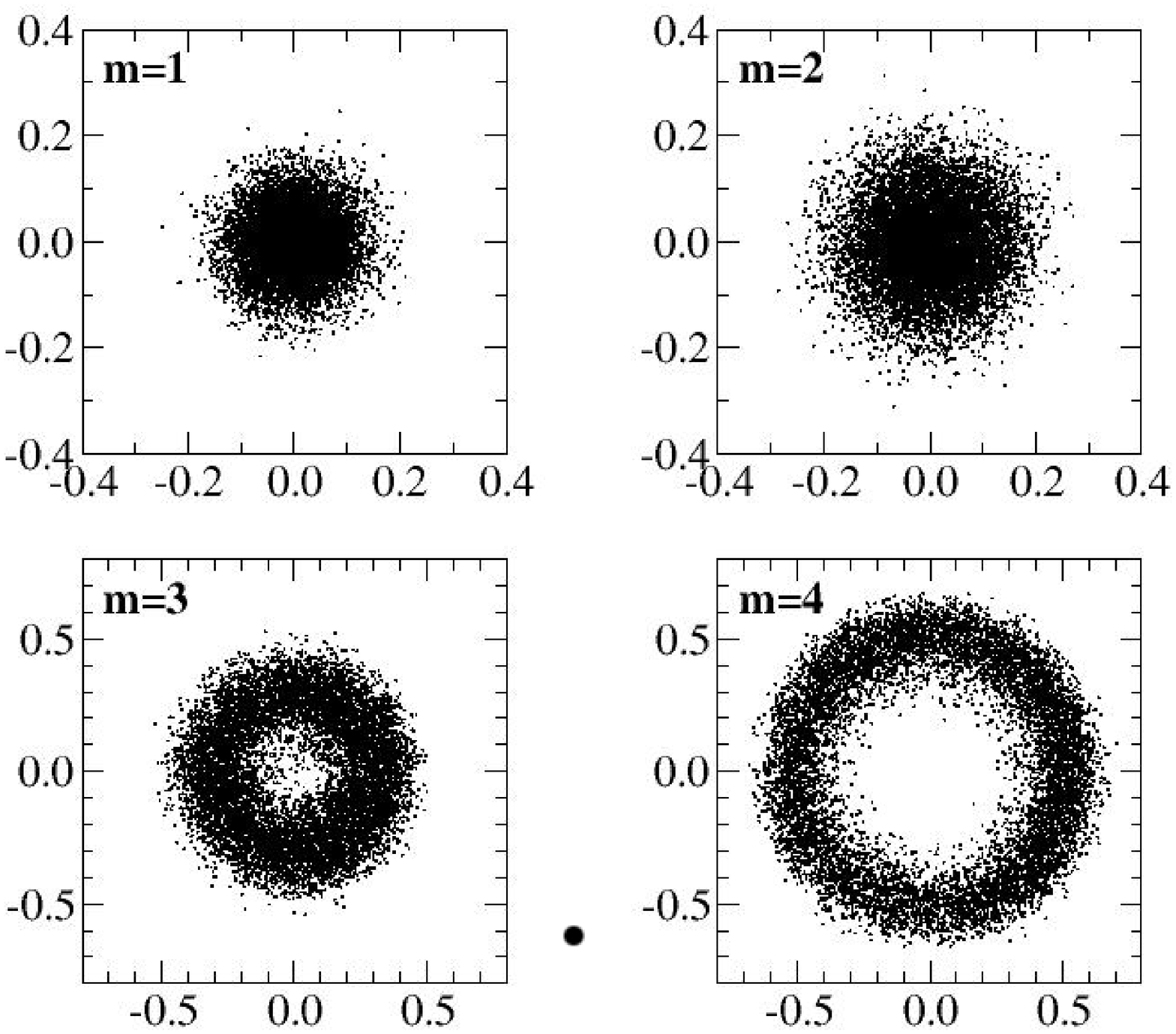}
\hspace{-1.3cm}

\includegraphics[width=0.75\textwidth,angle=0]
{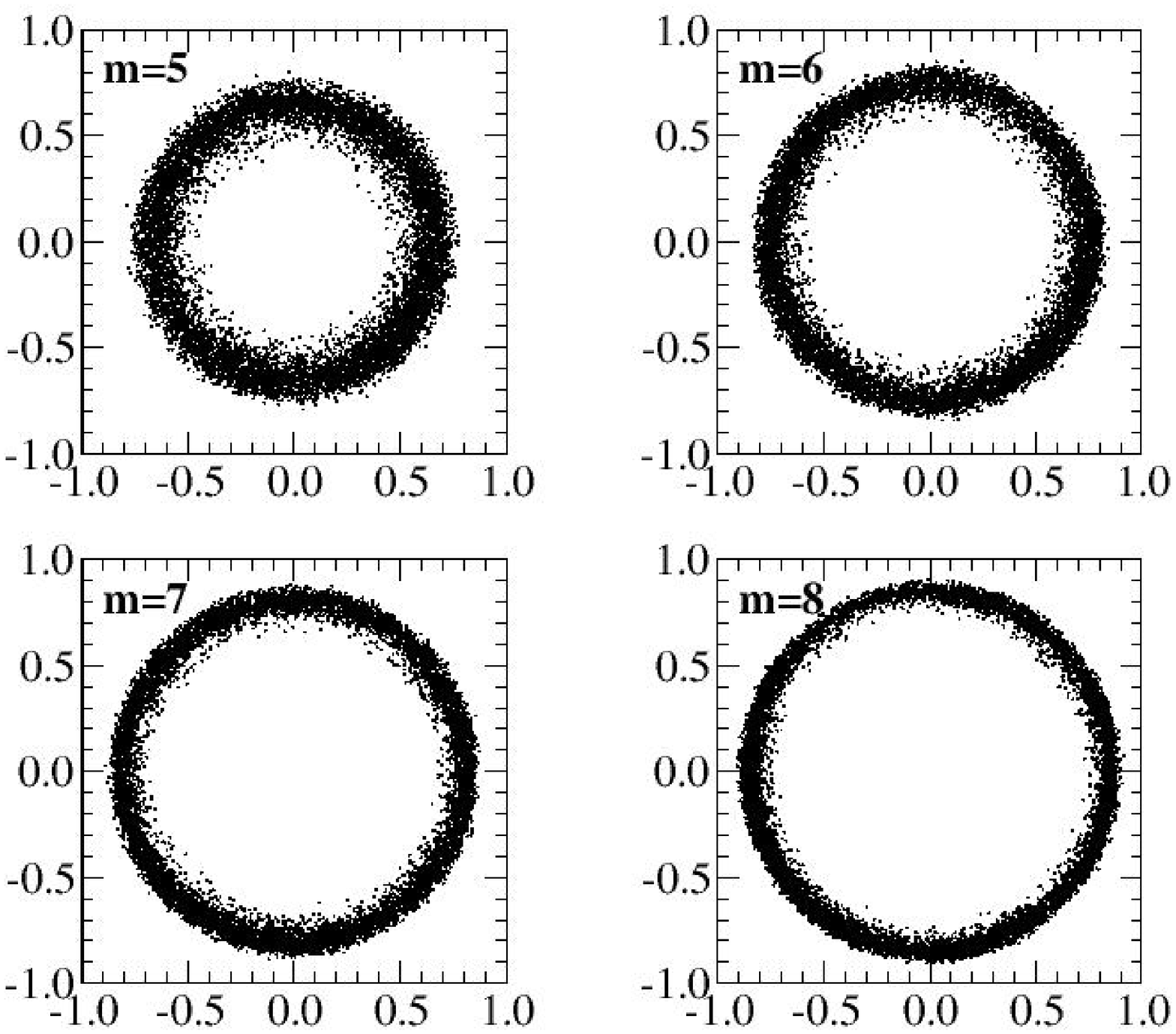}
\caption[]{Scatter plots of $\Pi_m$ (Im$\Pi_m$ versus Re$\Pi_m$) 
for 8 values of $m$ on a $12^2 \times 8$ 
lattice at $\beta=0.15$ (confined phase); the simulation parameters are 
$aM=0.05$, $\omega=0.1$, $\delta t=0.03$, MCR=50, FOM=1000, NMS=10000. For 
$m=1,2$ data are homogeneously distributed on a disc; for $m \geq 3$ data are 
homogeneously distributed on a ring whose radius increases with $m$.}
\label{plot_conf}
\end{figure}

\begin{figure}[htbp]
\centering
\hspace{-1.3cm}
\includegraphics[width=0.75\textwidth,angle=0]
{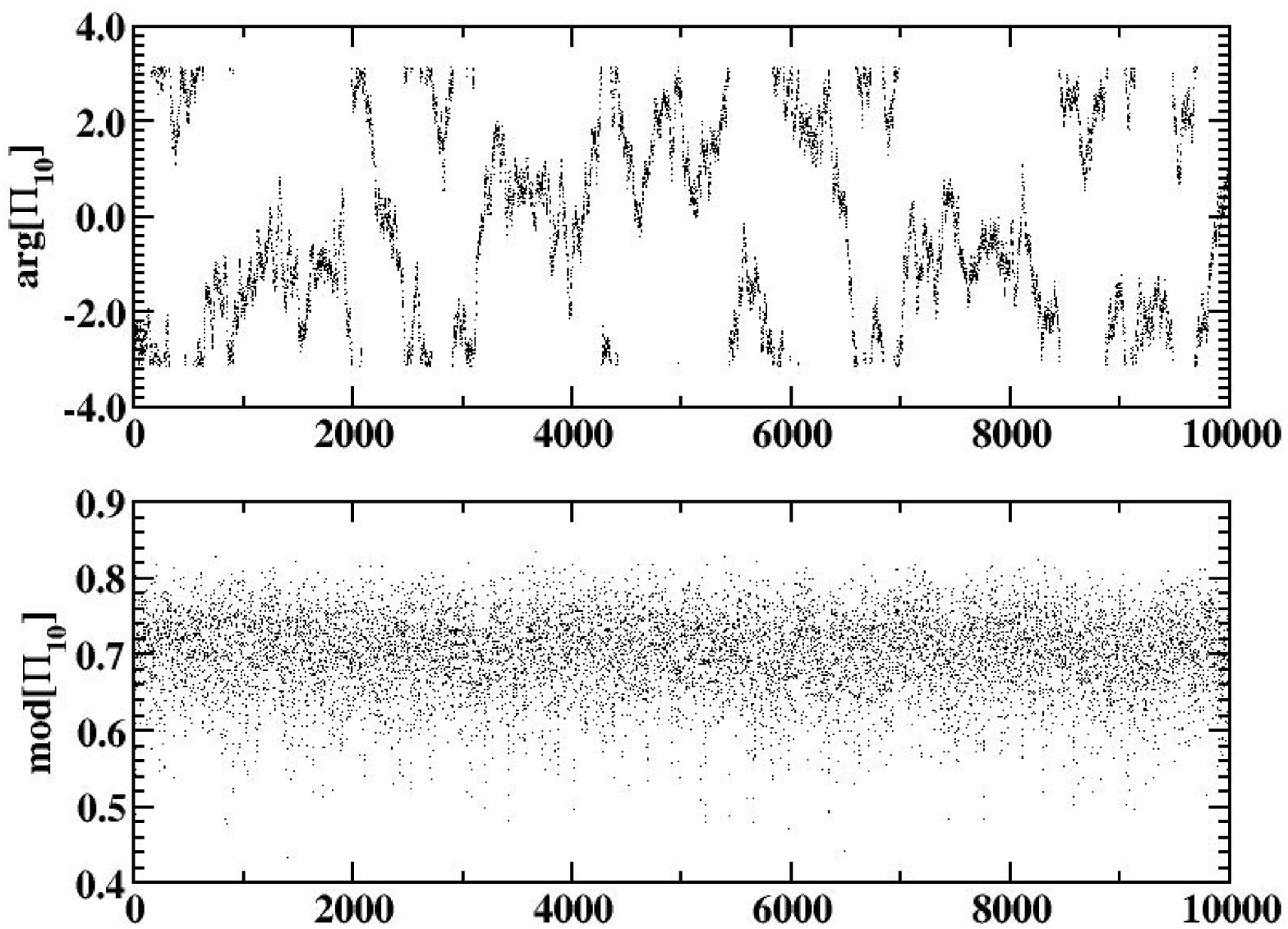}
\caption[]{Evolution with Monte Carlo time of modulus (down) and argument (up) 
of $\Pi_{10}$ on a $12^2 \times 8$ lattice at $\beta=0.05$ (confined phase); 
the simulation parameters are as in Figure~\ref{plot_conf}.}
\label{plot_conf_b}
\end{figure}

In Figure~\ref{plot_conf} we show scatter plots of the order parameter
$\Pi_m$ in the complex plane for a $\beta$ value known {\it a posteriori} 
to lie in the confined phase and for several values of $m$. 
For $m=1,2$ data distribute on a disk, while for 
$m \geq 3$ they are on a ring whose radius increases with $m$; in both 
cases data are uniformly spread around the origin.
Figure~\ref{plot_conf_b} gives the evolution in the simulation time of 
$\Pi_{10}$; it is interesting to observe that $\arg{\Pi_{10}}$ spans in a 
continuous way the interval $[0,2\pi]$ (with account of the $2\pi$
periodicity). 

\begin{figure}[htbp]
\centering
\hspace{-1.3cm}
\includegraphics[width=0.75\textwidth,angle=0]
{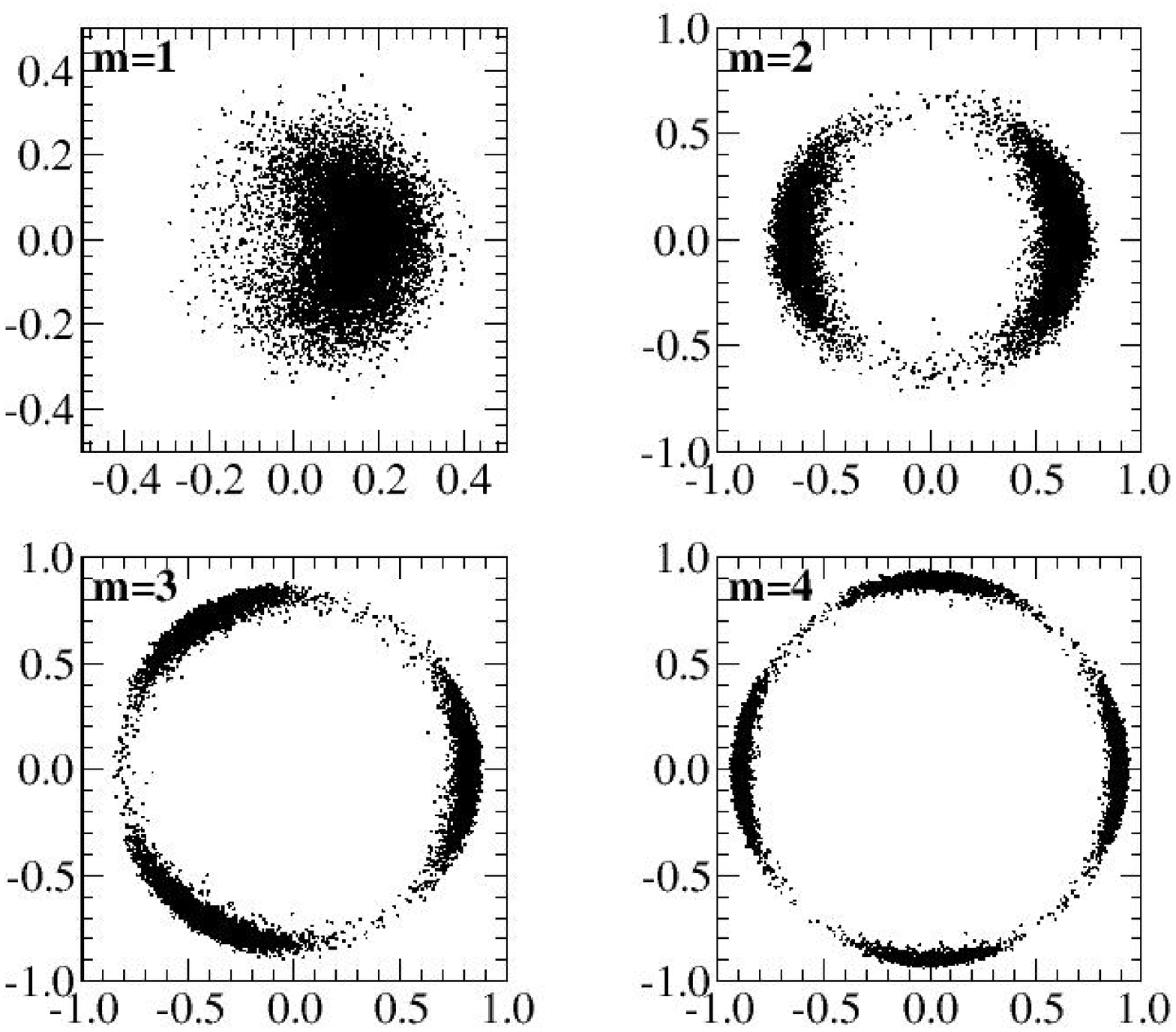}
\hspace{-1.3cm}

\includegraphics[width=0.75\textwidth,angle=0]
{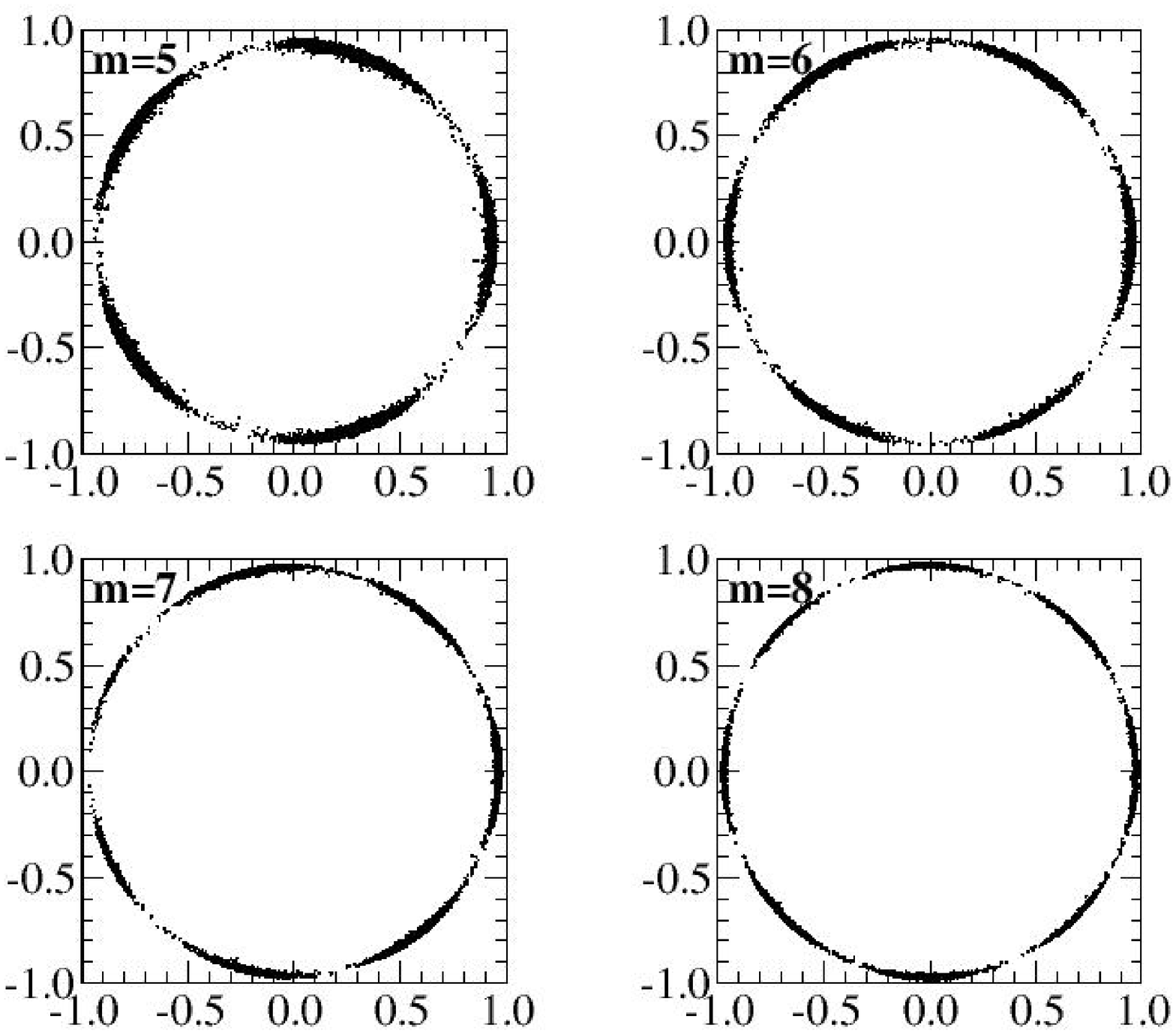}
\caption[]{Scatter plots of $\Pi_m$ (Im$\Pi_m$ versus Re$\Pi_m$) 
for 8 values of $m$ on a $12^2 \times 8$ 
lattice at $\beta=0.80$ (deconfined phase); the simulation parameters are 
$aM=0.05$, $\omega=0.1$, $\delta t=0.03$, MCR=50, FOM=1000, NMS=10000. In this 
case data are distributed on the $m$ roots of unity.}
\label{plot_deconf}
\end{figure}

\begin{figure}[htbp]
\centering
\hspace{-1.3cm}
\includegraphics[width=0.75\textwidth,angle=0]
{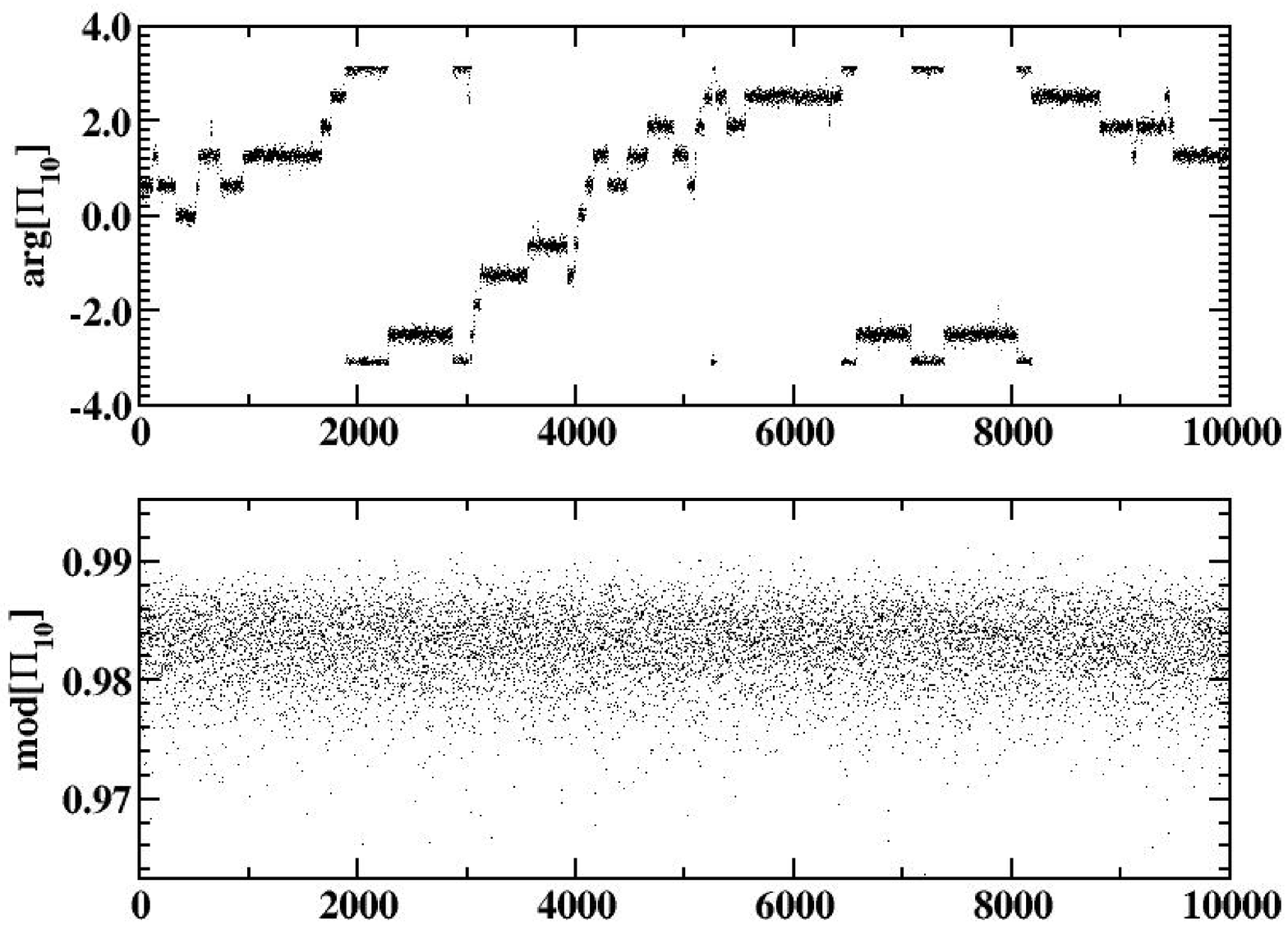}
\caption[]{Evolution with Monte Carlo time of modulus (down) and argument (up) 
of $\Pi_{10}$ on a $12^2 \times 8$ lattice at $\beta=0.80$ (deconfined phase); 
the simulation parameters are as in Figure~\ref{plot_deconf}.}
\label{plot_deconf_b}
\end{figure}

Figure~\ref{plot_deconf} is the same as Figure~\ref{plot_conf}, but 
for a value of $\beta$ in the deconfined phase. 
In this case the $\mathbb{Z}_m$ symmetry is broken and the values 
of the order parameter accumulate in correspondence of the $m$ roots of 
the identity. Accordingly, the evolution in simulation time of
$\arg{\Pi_{10}}$ spans the interval $[0,2\pi]$ in jerky way (see
Figure~\ref{plot_deconf_b}).

\begin{figure}[htbp]
\centering
\hspace{-1.3cm}
\includegraphics[width=0.75\textwidth,angle=0]
{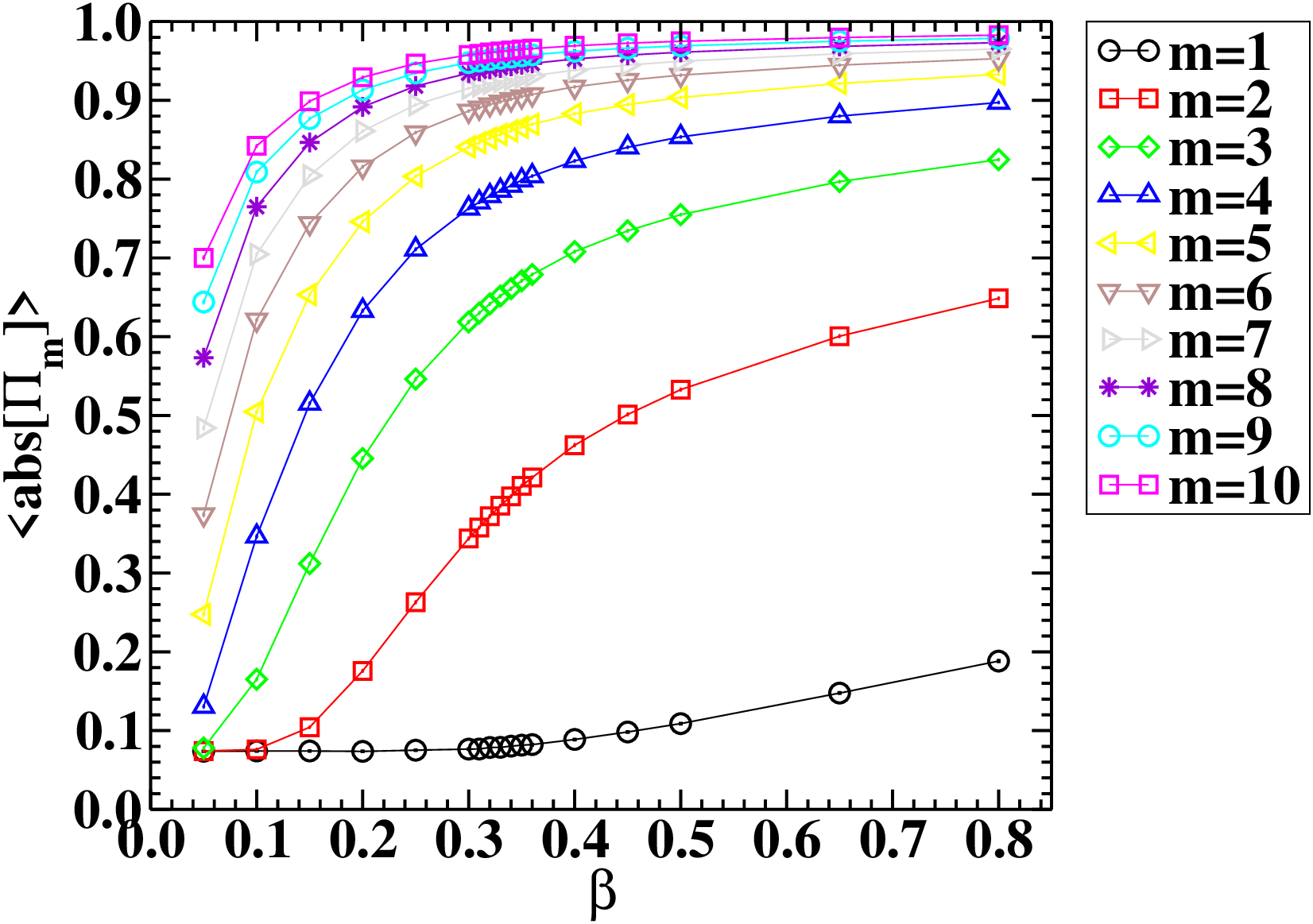}
\caption[]{Modulus of $\langle \Pi_m \rangle$ for 10 values of $m$ versus 
$\beta$ on a $12^2 \times 8$ lattice; the solid line is to guide the eye; the 
simulation parameters are $aM=0.05$, $\omega=0.1$, $\delta t=0.03$, MCR=50, 
FOM=1000, NMS=10000.}
\label{abs_pi_m_vs_beta}
\end{figure}

\begin{figure}[htbp]
\centering
\hspace{-1.3cm}
\includegraphics[width=0.75\textwidth,angle=0]
{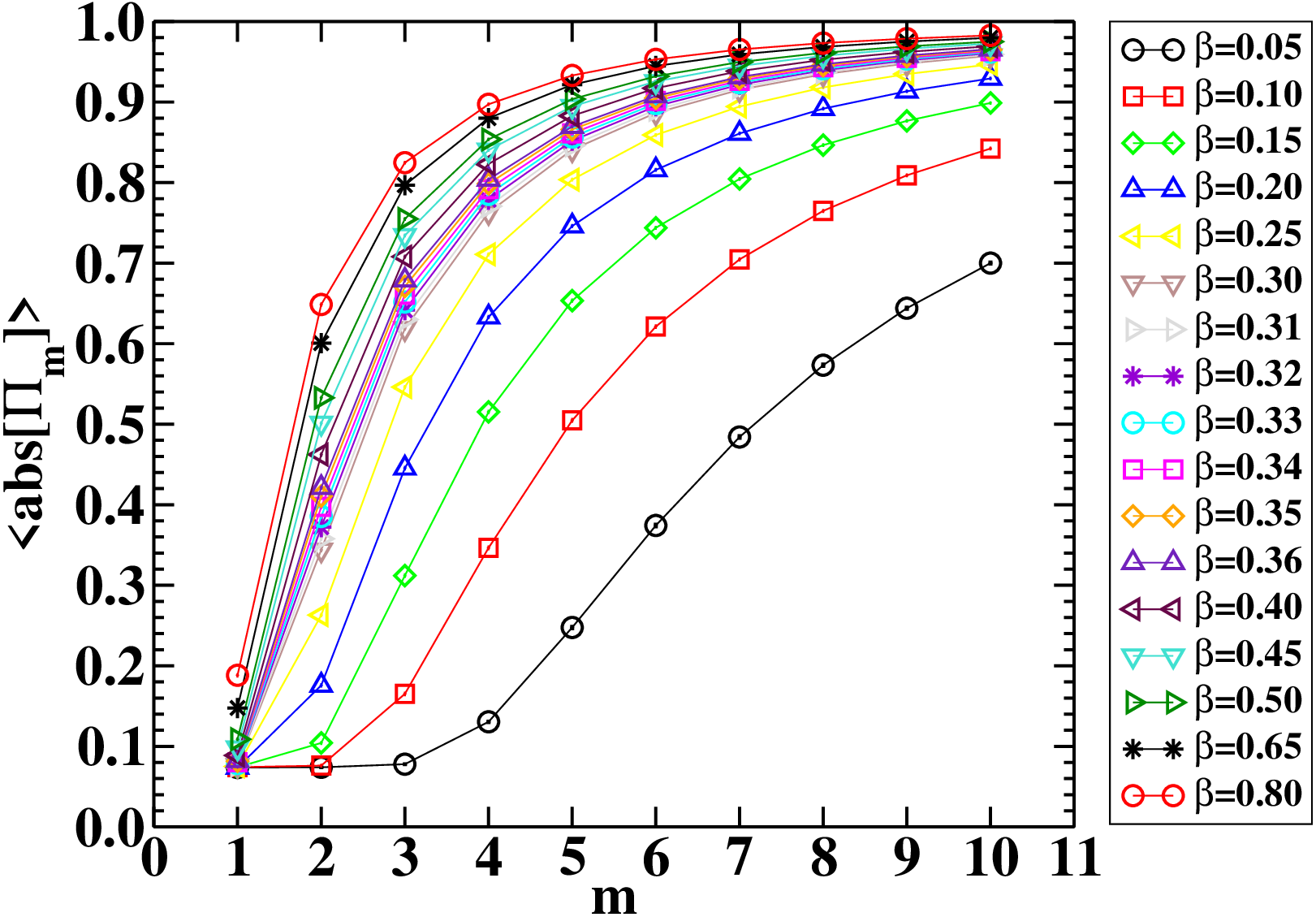}
\caption[]{Modulus of $\langle \Pi_m \rangle$ for various values of $\beta$ 
versus $m$ on a $12^2 \times 8$ lattice; the solid line is to guide the eye; 
the simulation parameters are as in Figure~\ref{abs_pi_m_vs_beta}.}
\label{abs_pi_m_vs_m}
\end{figure}

\begin{figure}[htbp]
\centering
\hspace{-1.3cm}
\includegraphics[width=0.75\textwidth,angle=0]
{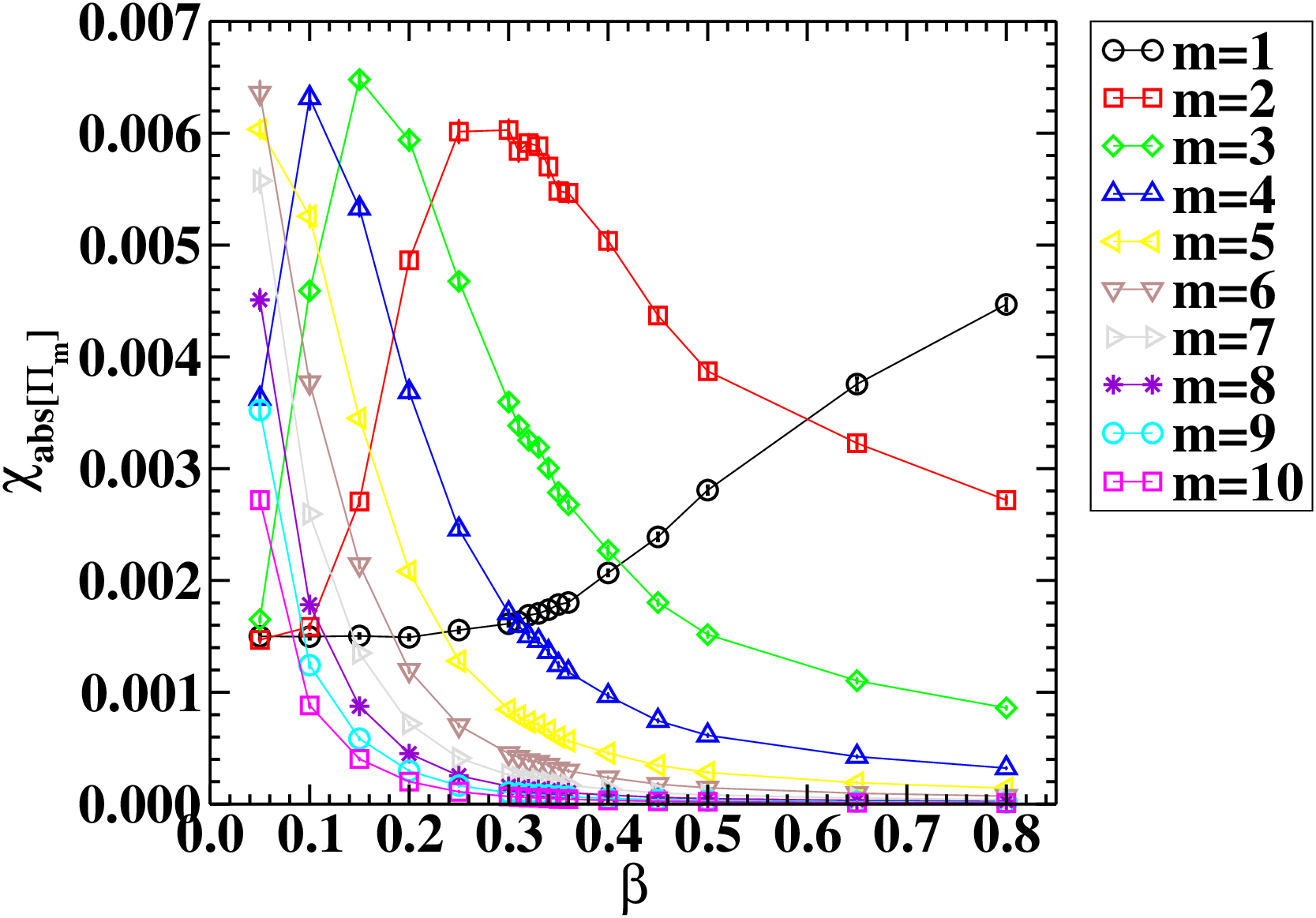}
\caption[]{Susceptibility of $\mbox{abs}[\Pi_m]$ for 10 values of $m$ versus 
$\beta$ on a $12^2 \times 8$ lattice; the solid line is to guide the eye; the 
simulation parameters are as in Figure~\ref{abs_pi_m_vs_beta}.}
\label{susc_pi_m}
\end{figure}

These scatter plots tell us that real and imaginary parts of $\langle\Pi_m
\rangle$ are zero in both phases, if, evidently, there is enough tunneling
in the deconfined phase and if the algorithm explores the whole configuration
space in the confined phase, as it seems to be the case on the lattice
$12^2\times 8$ used to obtain the above Figures. So, the more informative 
quantity here seems to be $\langle \mbox{abs}(\Pi_m) \rangle$.
The behavior of this quantity is shown in Figure~\ref{abs_pi_m_vs_beta} for 
varying $\beta$ and 
for several values of $m$ and in Figure~\ref{abs_pi_m_vs_m} for varying $m$ and
for different values of $\beta$. Figure~\ref{abs_pi_m_vs_beta} shows that 
there is a smooth increase with $\beta$, more evident for higher values of 
$m$. Data do not allow to clearly single out any transition; indeed, also
the susceptibility of $\mbox{abs}(\Pi_m)$ does not show clear peaks for varying
$\beta$ in the same region (see Figure ~\ref{susc_pi_m}). Therefore, we can
conclude that this operator is not convenient to distinguish the two
phases which lead to the different behaviors shown in Figures~\ref{plot_conf} 
and ~\ref{plot_deconf}.

This conclusion suggests to move to the operator $\Pi_m^m$, which evidently 
has no effect in the angular distribution in the confined phase, i.e. in the
situation of Figure~\ref{plot_conf}, while ``concentrates'' the angular
distribution along the first of the $m$ roots of the identity in the
deconfined phase, i.e. in the situation of Figure~\ref{plot_deconf}. 
Equivalently, we can say that $\langle \arg{\Pi_m^m}\rangle=0$ in the
deconfined phase. The operator $\Pi_m^m$ should therefore permit to 
distinguish between broken and unbroken symmetry phases; since 
$\mbox{Im}\langle \Pi_m^m \rangle$ is always zero, we consider from the 
beginning $\mbox{Re}\langle \Pi_m^m \rangle$.

\begin{figure}[htbp]
\centering
\hspace{-1.3cm}
\includegraphics[width=0.75\textwidth,angle=0]
{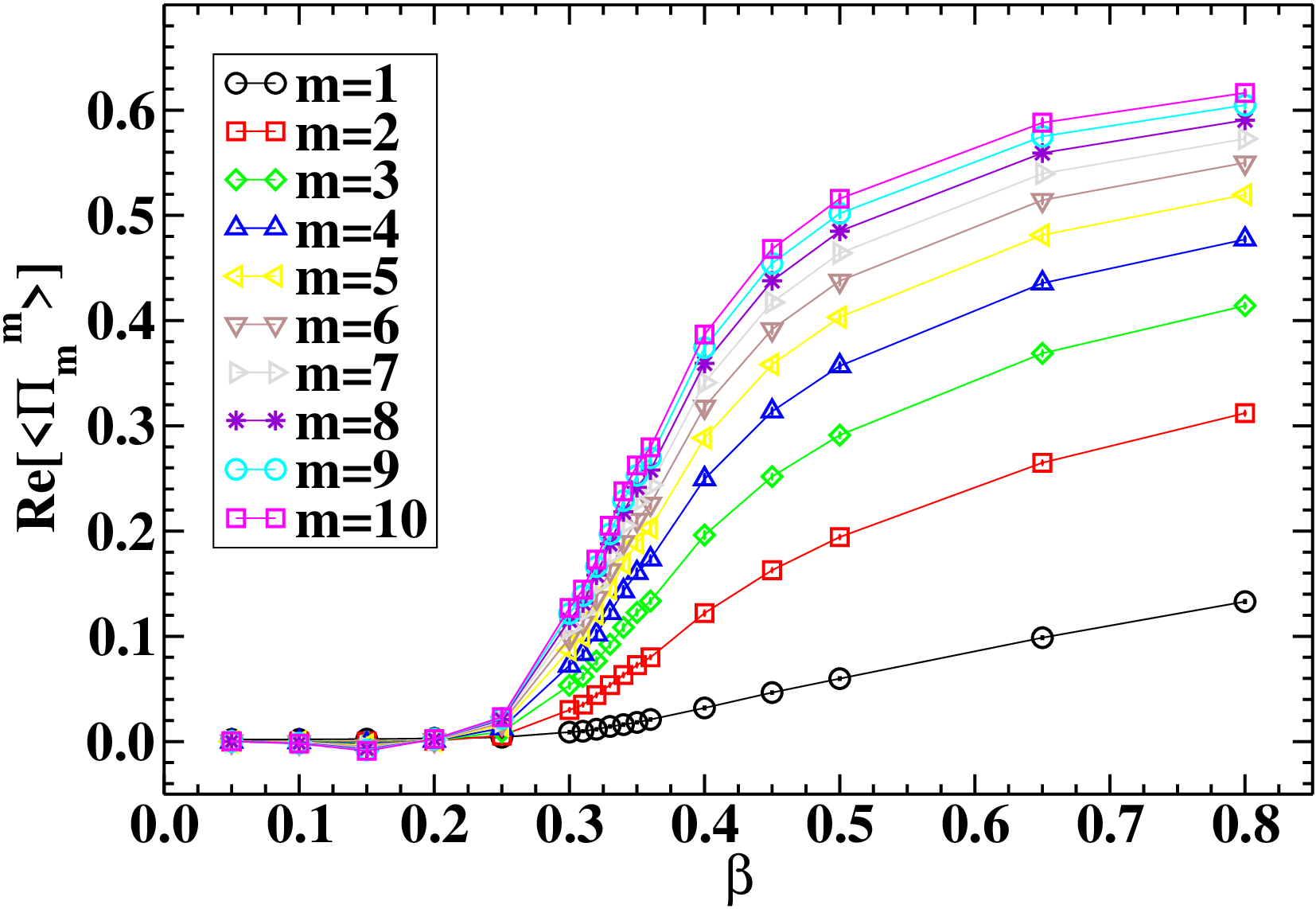}
\caption[]{Real part of $\langle \Pi^m_m \rangle$ for 10 values of $m$ versus 
$\beta$ on a $12^2 \times 8$ lattice; the solid line is to guide the eye; the 
simulation parameters are $aM=0.05$, $\omega=0.1$, $\delta t=0.03$, MCR=50, 
FOM=1000, NMS=10000.}
\label{re_pi_m_m_vs_beta}
\end{figure}

\begin{figure}[htbp]
\centering
\hspace{-1.3cm}
\includegraphics[width=0.75\textwidth,angle=0]
{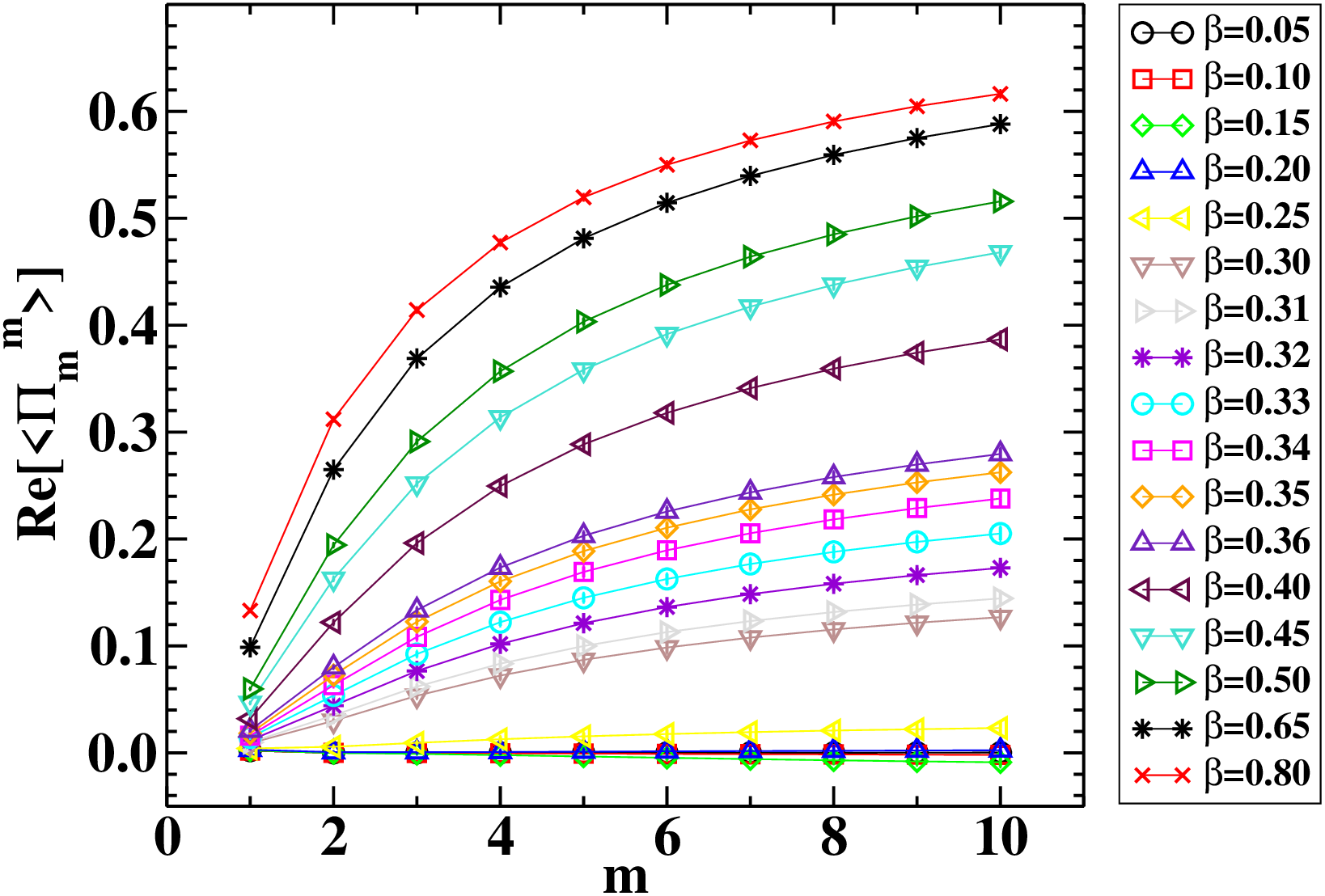}
\caption[]{Real part of $\langle \Pi^m_m \rangle$ for various values of 
$\beta$ versus $m$ on a $12^2 \times 8$ lattice; the solid line is to guide 
the eye; the simulation parameters are as in Figure~\ref{re_pi_m_m_vs_beta}.}
\label{re_pi_m_m_vs_m}
\end{figure}

\begin{figure}[htbp]
\centering
\hspace{-1.3cm}
\includegraphics[width=0.75\textwidth,angle=0]
{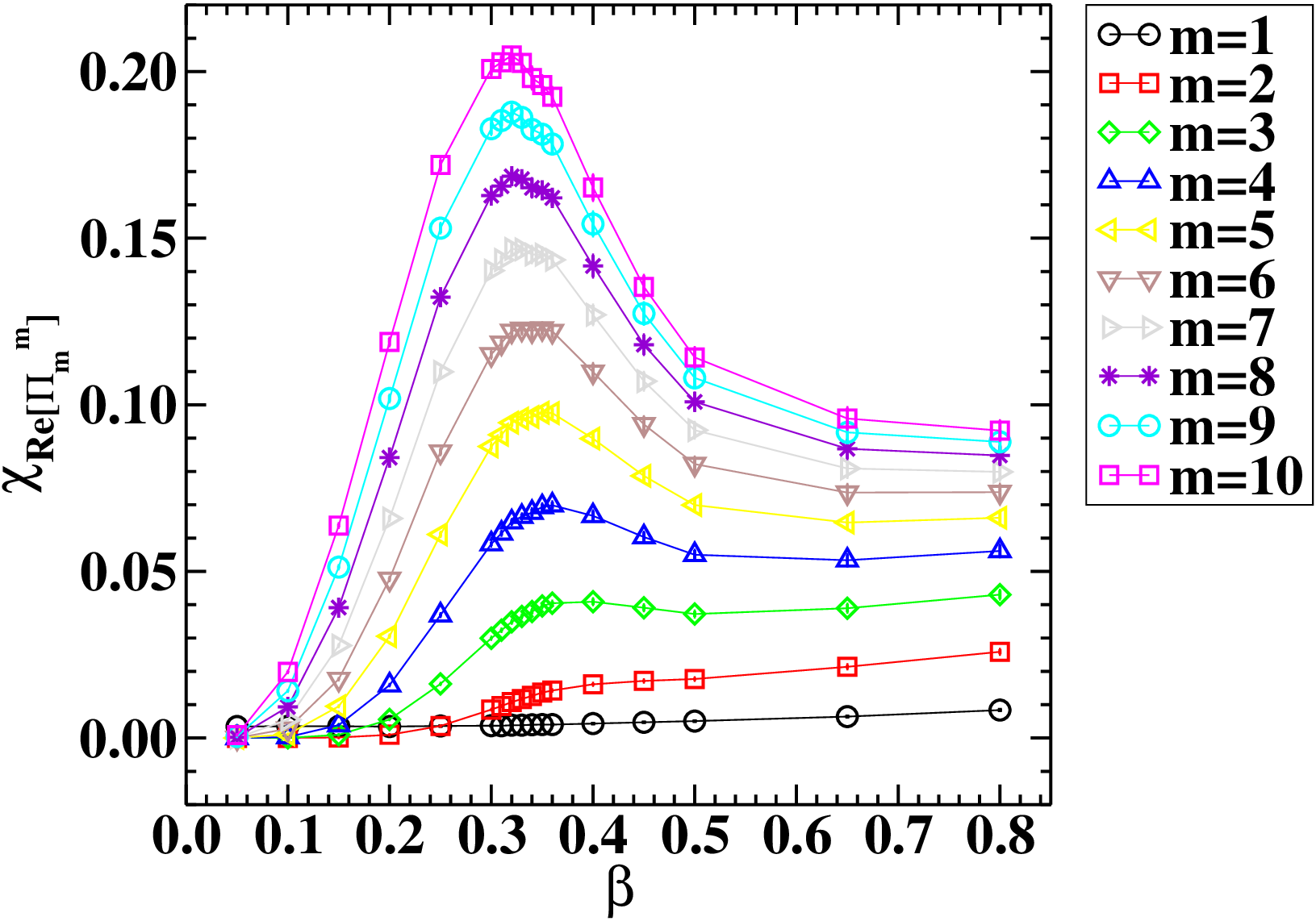}
\caption[]{Susceptibility of $\mbox{Re}[\Pi^m_m]$ for 10 values of $m$ versus 
$\beta$ on a $12^2 \times 8$ lattice;  the solid line is to guide the eye; the 
simulation parameters are as in Figure~\ref{re_pi_m_m_vs_beta}.}
\label{susc_re_pi_m_m}
\end{figure}

Results are shown in Figures~\ref{re_pi_m_m_vs_beta} and ~\ref{re_pi_m_m_vs_m}.
In this case we have a clear hint on the position of the transition point
between the two phases. In particular, Figure~\ref{re_pi_m_m_vs_beta} indicates
that, for all values of $m>1$, there is a sharp increase of the 
signal at $\beta\simeq 0.3$, while Figure~\ref{re_pi_m_m_vs_m} suggests
that the increase is more pronounced the higher is $m$. For $m=1$ 
data show no transition at all, as it must be, since $m=1$ means integer
electric charge. This is confirmed in Figure ~\ref{susc_re_pi_m_m} where a peak
in the susceptibility of the operator $\mbox{Re} \Pi_m^m$ 
appears for all values of $m>1$ at a $\beta_c$ value which, on this
lattice, has a mild dependence on $m$, namely, it decreases with $m$ 
and seems to stabilize for high $m$ around $\beta_c=0.32$.

\begin{figure}[htbp]
\centering
\hspace{-1.3cm}
\includegraphics[width=0.75\textwidth,angle=0]
{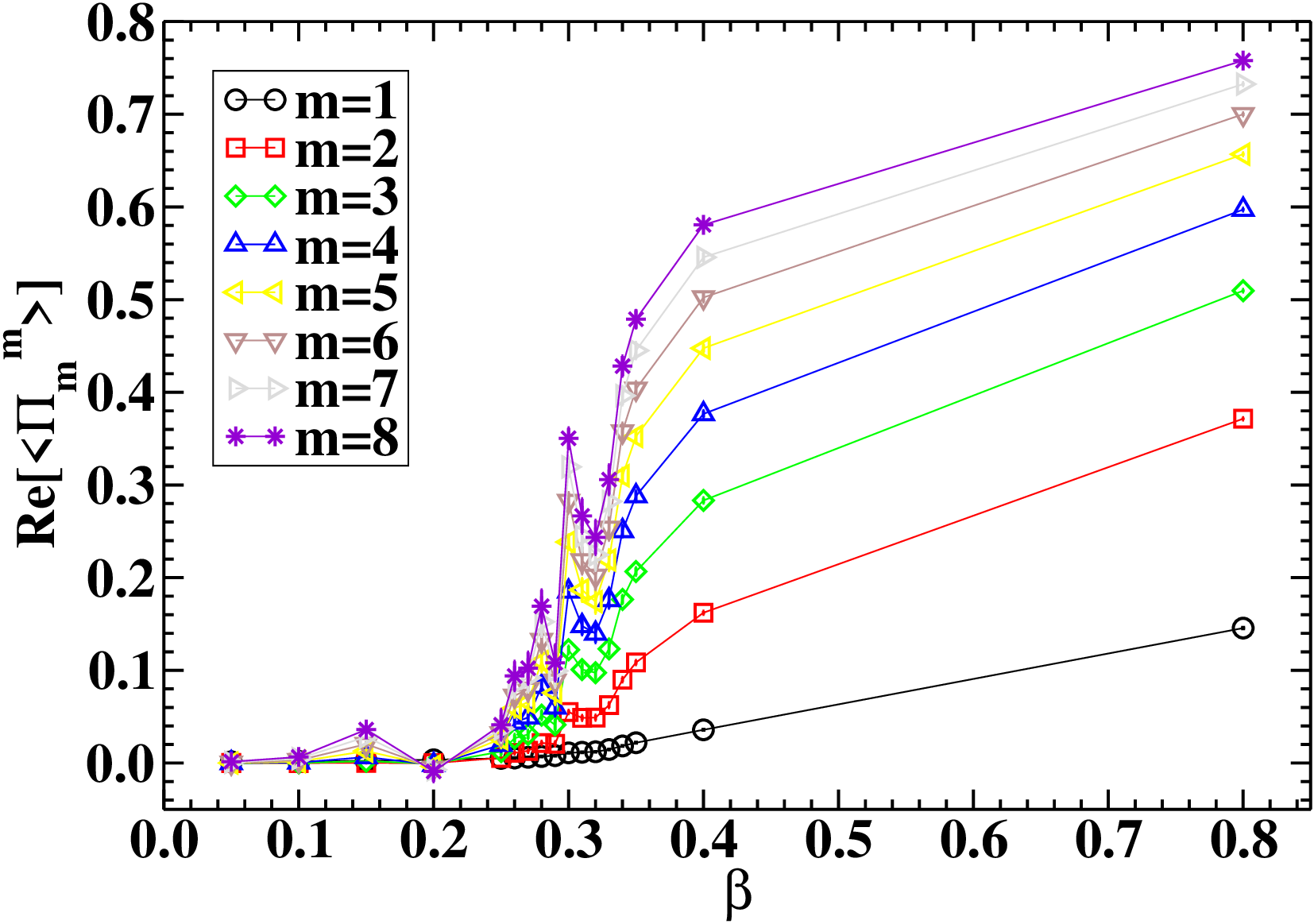}
\caption[]{Real part of $\langle \Pi^m_m \rangle$ for 8 values of $m$ versus 
$\beta$ on a $32^2 \times 8$ lattice;  the solid line is to guide the eye; the 
simulation parameters are $aM=0.05$, $\omega=0.2$, $\delta t=0.03$, MCR=10, 
FOM=10, NMS=100000.}
\label{re_pi_m_m_vs_beta_32}
\end{figure}

\begin{figure}[htbp]
\centering
\hspace{-1.3cm}
\includegraphics[width=0.75\textwidth,angle=0]
{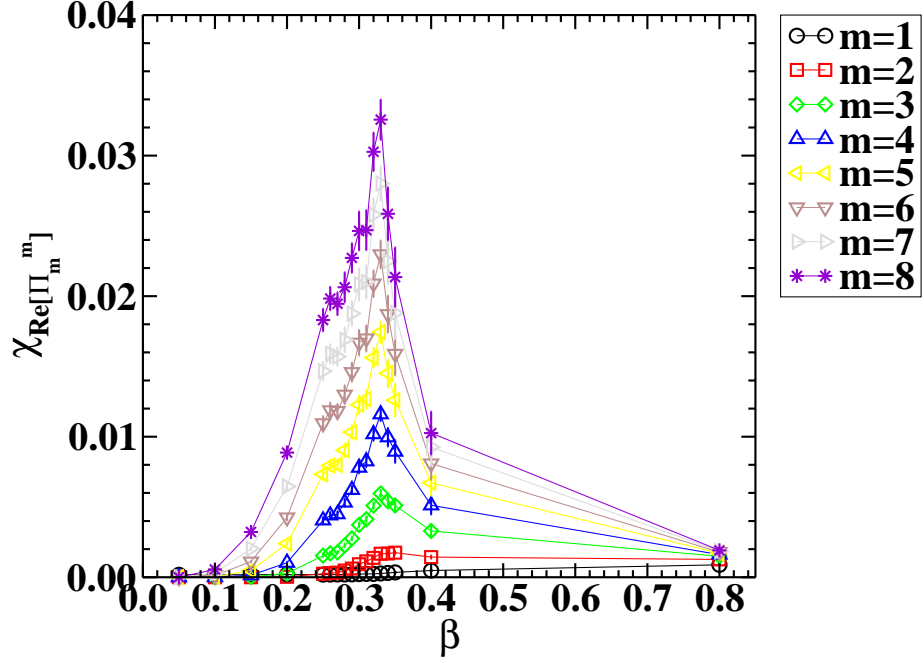}
\caption[]{Susceptibility of $\mbox{Re}[\Pi^m_m]$ for 8 values of $m$ versus 
$\beta$ on a $32^2 \times 8$ lattice;  the solid line is to guide the eye; the 
simulation parameters are as in Figure~\ref{re_pi_m_m_vs_beta_32}.}
\label{susc_re_pi_m_m_32}
\end{figure}

In Figures~\ref{re_pi_m_m_vs_beta_32} and ~\ref{susc_re_pi_m_m_32} we show the 
same results obtained on a lattice $32^2 \times 8$. In this case the values 
of $\beta$ where the peaks appear are more stable with $m$ and we can estimate 
the critical value as $\beta_c=0.33$. This leads us to argue that the
$m$-dependence of $\beta_c$ seen on the smaller lattice could be a finite 
volume effect.

It is important to compare the numerical result for $\beta_c$ with the
theoretical expectation discussed in Section~\ref{subsec:phase}, in particular
with the value of $\beta_c$ given in Eq.~\ref{betac}: 
$\beta_c^{th}=\pi^{-1} \simeq 0.318\ldots$. This value is impressively
close to our numerical result, thus supporting the theoretical scenario
proposed in Ref.~\cite{Grignani:1995iv} and the conjectured BKT transition 
(see also Ref.~\cite{Kovner:1992}).

\begin{figure}[htbp]
\centering
\hspace{-1.3cm}
\includegraphics[width=0.75\textwidth,angle=0]
{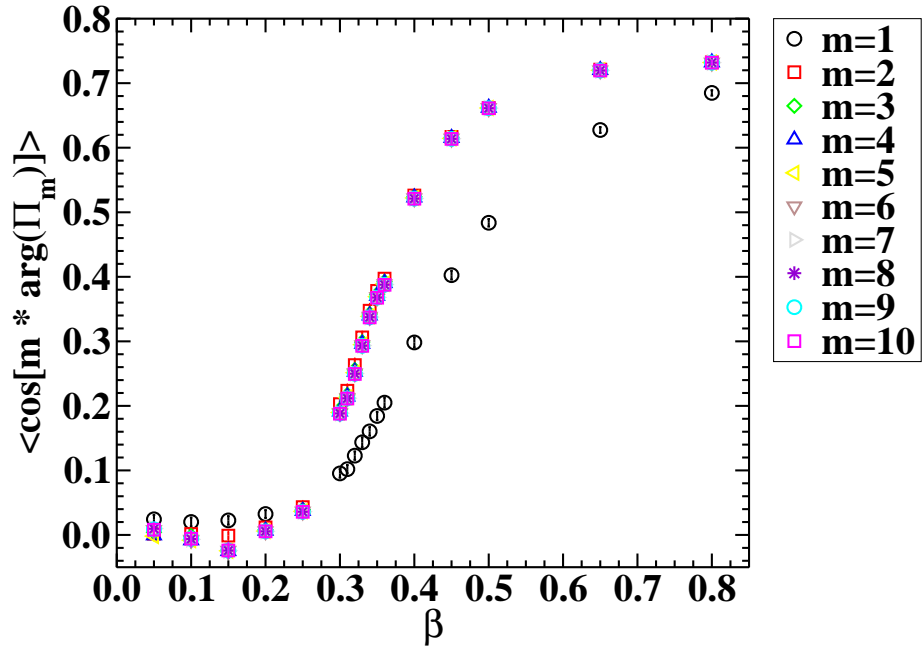}
\caption[]{$\langle \cos [ m \times \arg \Pi_m ] \rangle$ for 10 values of $m$ 
versus $\beta$ on a $12^2 \times 8$ lattice; the simulation parameters are 
$aM=0.05$, $\omega=0.1$, $\delta t=0.03$, MCR=50, FOM=1000, NMS=10000.}
\label{cos_vs_beta}
\end{figure}

\begin{figure}[htbp]
\centering
\hspace{-1.3cm}
\includegraphics[width=0.75\textwidth,angle=0]
{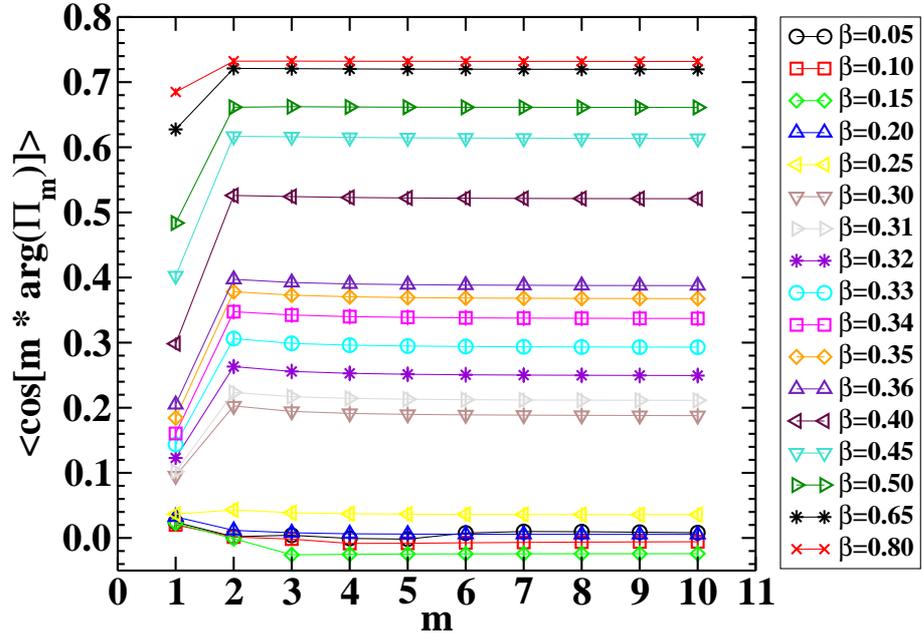}
\caption[]{$\langle \cos [ m \times \arg \Pi_m ] \rangle$ for various values 
of $\beta$ versus $m$ on a $12^2 \times 8$ lattice;  the solid line is to guide
the eye; the simulation parameters are as in Figure~\ref{cos_vs_beta}.}
\label{cos_vs_m}
\end{figure}

\begin{figure}[htbp]
\centering
\hspace{-1.3cm}
\includegraphics[width=0.75\textwidth,angle=0]
{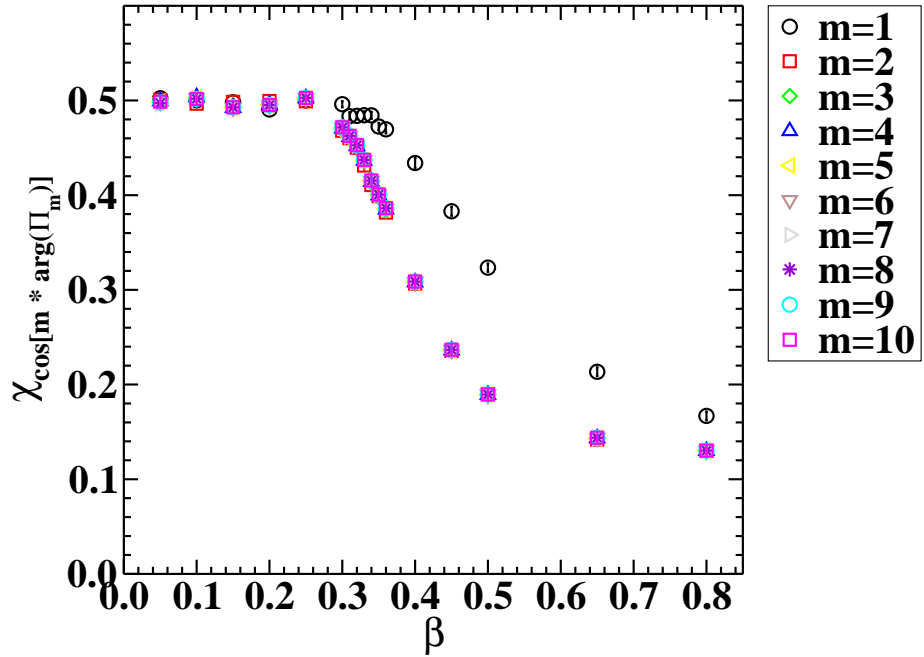}
\caption[]{Susceptibility of $\cos [ m \times \arg \Pi_m ]$ for 10 values of 
$m$ versus $\beta$ on a $12^2 \times 8$ lattice; the simulation parameters are 
as in Figure~\ref{cos_vs_beta}.}
\label{susc_cos}
\end{figure}

In Figures~\ref{cos_vs_beta} and ~\ref{cos_vs_m} we consider the observable  
$\langle \cos [ m \times \arg \Pi_m ] \rangle$, which should give similar
information as the previous observable, concerning the location of the
transition point. We see that data for $m>1$ fall on top of each other, 
while data for $m=1$ describe a different curve, but also exhibit a sharp 
increase at the same point as data for $m>1$. Since we know that for 
$m=1$ the observable cannot be an order parameter, this effect should 
be an artifact of the finite volume. Unfortunately, the susceptibility, 
shown in Figure~\ref{susc_cos}, does not have a peak structure, but shows 
rather a jump between two constant values, therefore it does not allow 
for an accurate determination of $\beta_c$.

\subsection{Polyakov loop correlators}
\label{subsec:correl}

We have studied the wall-wall correlation between two Polyakov 
loops\footnote{Note that the value of $G(r)=\mbox{Im} \langle \Pi(0)\Pi^*(r) 
\rangle$ is always compatible with zero.}
\begin{equation}
G(r)=\mbox{Re} \langle \Pi(0)\Pi^*(r) \rangle
\end{equation}
for two values of $\beta$, $\beta=0.25$ and $\beta=0.40$, the first below the 
critical temperature and the other above it.

It is worth noting here that the dimensionless parameter $g^2/M$ is equal
to 80 for $\beta=0.25$ and to 50 for $\beta=0.40$. The other important 
dimensionless parameter $T/M$ is equal to 2.5 in all simulations performed
in this work. These values for $g^2/M$ and $T/M$ are far from the large 
$M$ limit, where the theoretical analysis of 
Refs.~\cite{Grignani:1995iv,Grignani:1995hx} was carried on (see 
Subsection~\ref{subsec:phase}). However, the authors of 
Ref.~\cite{Grignani:1995hx} state that they expect their results 
are stable toward lower electron masses.

\begin{figure}[htbp]
\centering
\hspace{-1.3cm}
\includegraphics[width=0.65\textwidth,angle=0]
{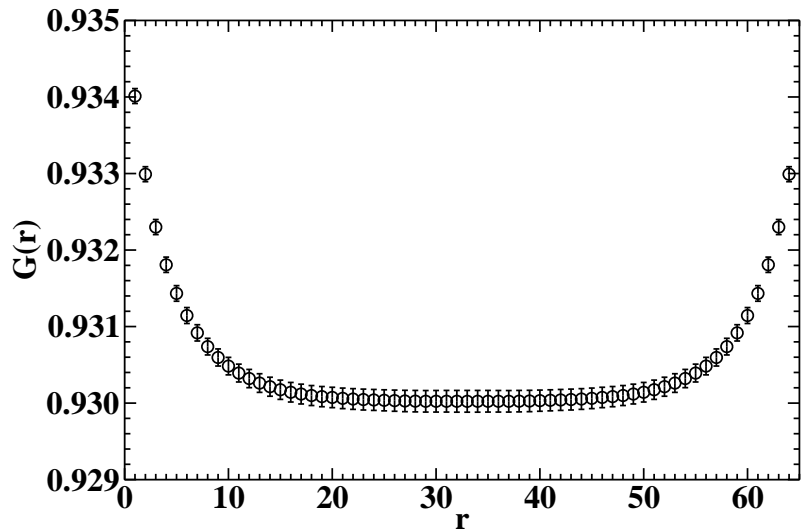}
\caption[]{$G(r)$ versus $r$ on a $64^2 \times 8$ lattice at $\beta=0.40$ 
(deconfined phase); the simulation parameters are $m=10$, $aM=0.05$, 
$\omega=0.1$, $\delta t=0.03$, MCR=50, FOM=250, NMS=8000.}
\label{G_beta_0.40}
\end{figure}

\begin{figure}[htbp]
\centering
\hspace{-1.3cm}
\includegraphics[width=0.75\textwidth,angle=0]
{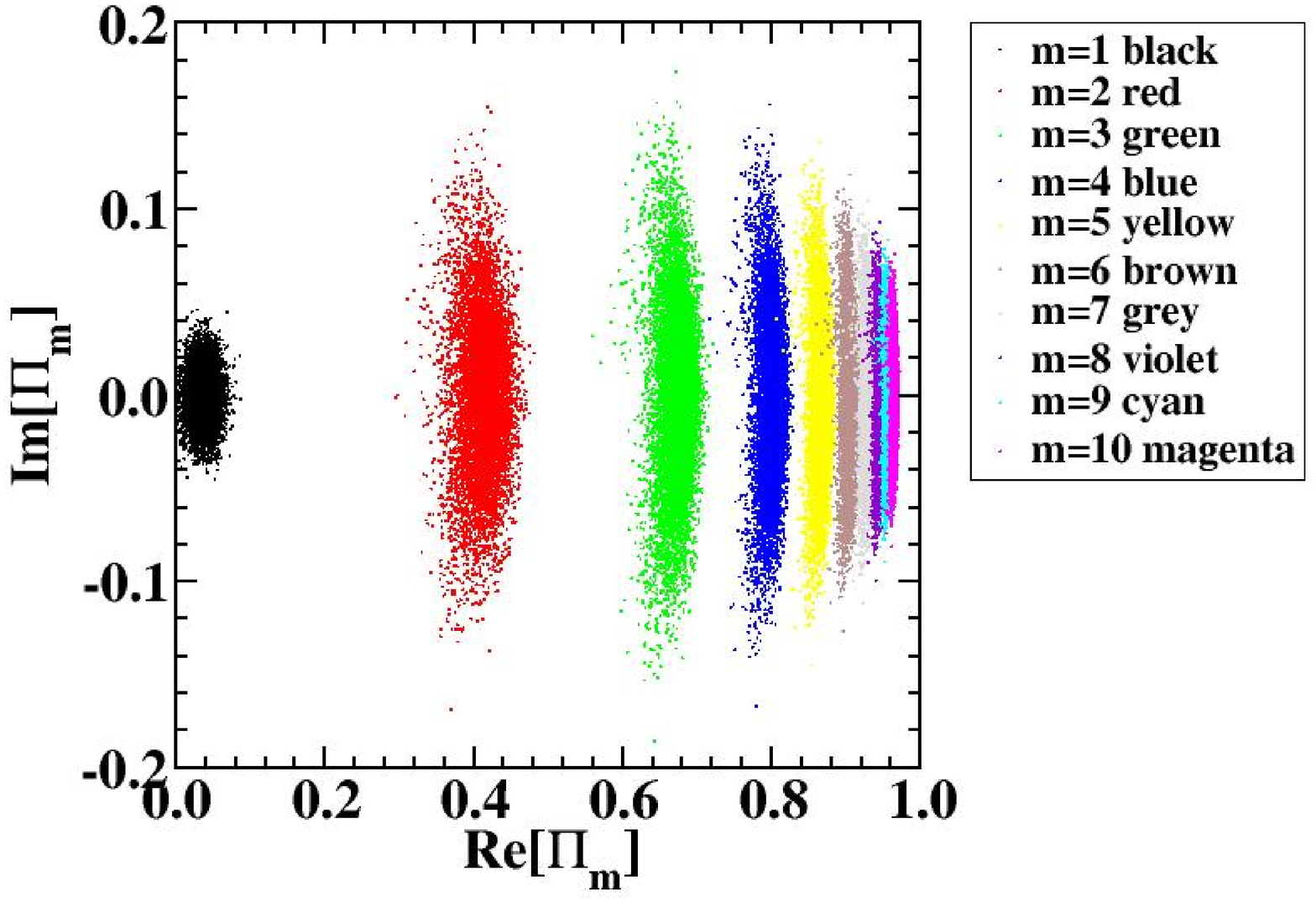}
\caption[]{Scatter plots of $\Pi_m$ for 10 values of $m$ on on a $64^2 \times 
8$ lattice at $\beta=0.40$ (deconfined phase); the simulation parameters are
$aM=0.05$, $\omega=0.1$, $\delta t=0.03$, MCR=50, FOM=250, NMS=8000.}
\label{scatterplot_0.40}
\end{figure}

For $\beta=0.40$ we have simulated the system on two different lattices: 
$32^2 \times 8$ with $\omega=0.2$, MCR=10, FOM=10,  NMS=100000;
$64^2 \times 8$ with $\omega=0.1$, MCR=50, FOM=250, NMS=8000.
Data for the correlator, shown in Figure~\ref{G_beta_0.40} (see also
Figure~\ref{scatterplot_0.40} for the scatter plot of $\Pi_m$) have been 
fitted, in a range $[r_{min},r_{max}]$, using the law
\beq
G(r)=A \left( e^{-{\cal M} r}+e^{-{\cal M} (N_\sigma-r)} \right)+C\;;
\label{eq-g1}
\eeq
for both lattices and for the different values of $m$ we have obtained
$\chi^2$/d.o.f. $<1$. 
We have done ``sliding window'' fits, that is we have varied the value 
of $r_{min}$ and $r_{max}$ until obtaining stable values for the fit 
parameters.

The most important result is that $a{\cal M}$ depends neither on the value of 
$m$, and so on the fractional charge, nor on the lattice size: 
$a{\cal M}=0.20 \pm 0.05$. It is instructive to determine the
``correlation length'' related to this mass: $\xi/a\equiv1/(a{\cal M})=
5.00 \pm 1.25$; this value, much smaller than the lattice size $\xi/a
\ll N_\sigma$, justifies {\it a posteriori} the fit with an exponential 
function. We can conclude that this value has a physical meaning: it is the 
Debye mass characterizing the phase above the critical temperature.

Also for $\beta=0.25$ we have studied the system on two lattices: 
$32^2 \times 8$ with $\omega=0.1$, MCR=50, FOM=1000, NMS=10000, and
$64^2 \times 8$ with $\omega=0.1$, MCR=50, FOM=250,  NMS=10000.

\begin{figure}[htbp]
\centering
\hspace{-1.3cm}
\includegraphics[width=0.65\textwidth,angle=0]
{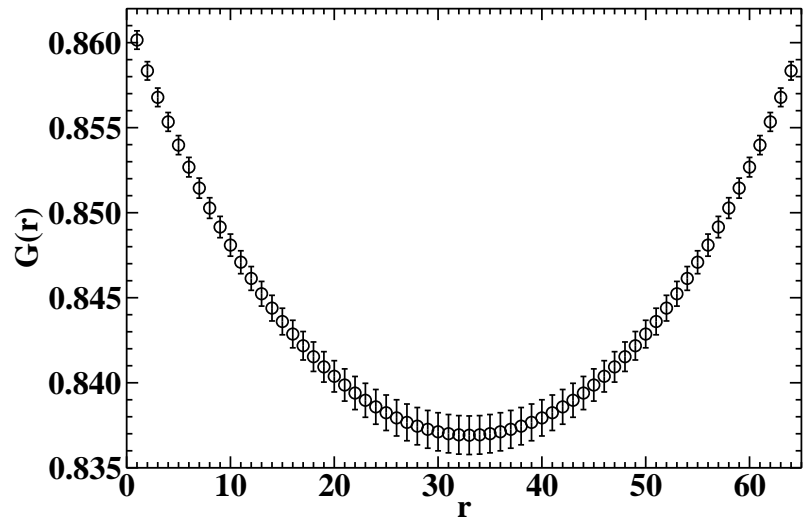}
\caption[]{$G(r)$ versus $r$ on a $64^2 \times 8$ lattice at $\beta=0.25$ 
(confined phase); the simulation parameters are $m=10$, $aM=0.05$, 
$\omega=0.1$, $\delta t=0.03$, MCR=50, FOM=250, NMS=10000.}
\label{G_beta_0.25}
\end{figure}

\begin{figure}[htbp]
\centering
\hspace{-1.3cm}
\includegraphics[width=0.75\textwidth,angle=0]
{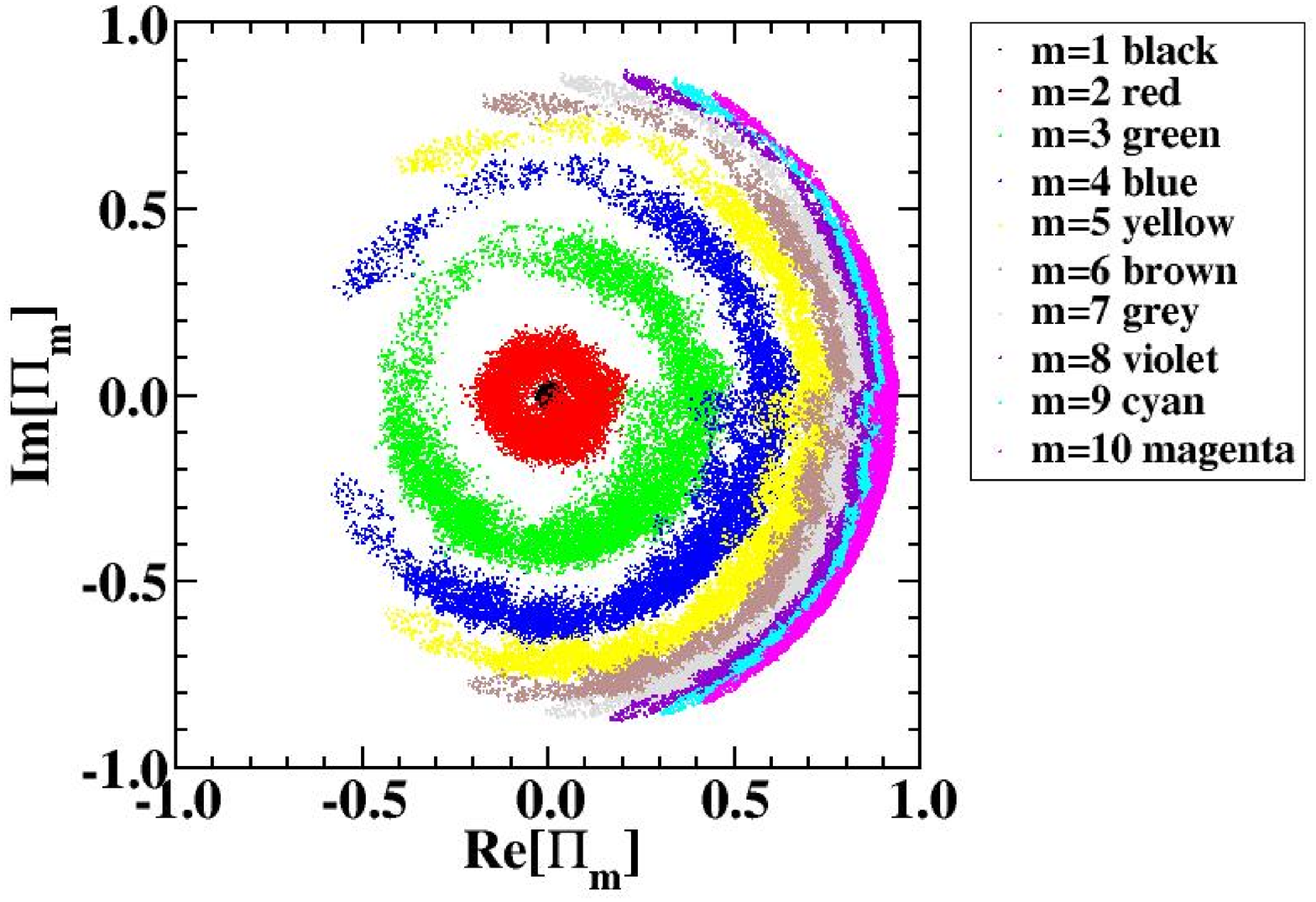}
\caption[]{Scatter plots of $\Pi_m$ for 10 values of $m$ on on a $64^2 \times 
8$ lattice at $\beta=0.25$ (confined phase); the simulation parameters are
$aM=0.05$, $\omega=0.1$, $\delta t=0.03$, MCR=50, FOM=250, NMS=10000.}
\label{scatterplot_0.25}
\end{figure}

As discussed in Sec.~\ref{subsec:phase}, we expect that here the best fit 
to the correlator should be given by a power law (see Eq.~(\ref{correl_0}));
in any case, to be conservative, we consider both the exponential and the 
power law. Therefore we should interpolate data shown in  
Figure~\ref{G_beta_0.25} with the following two functions:
\beq
G(r)=A \left( e^{-{\cal M} r}+e^{-{\cal M} (N_\sigma-r)} \right)+C\;,
\label{fit_exp}
\eeq
\beq
G(r)=A \left( r^{-\eta}+ (N_\sigma-r)^{-\eta} \right)+C\;.
\label{fit_pot}
\eeq
Here the constant $C$ should zero, since the $\mathbb{Z}$ symmetry is 
unbroken, however, we included it to take into account the
possibility that, due to not enough large statistics, the whole 
configuration space is not explored by the simulation algorithm, this
leading to $\langle \Pi_m\rangle \neq 0$ (see Figure~\ref{scatterplot_0.25}).
We note, however, that the large distance correlator does not go to zero
(see Figure~\ref{G_beta_0.25}); this happens because of the small values 
of ${\cal M}$ and $\eta$. In this case, if we try to fit data 
using directly Eqs.~(\ref{fit_exp}) or~(\ref{fit_pot}), we find that the value 
of $C$ is not equal to the expected value, i.e. 
$\mbox{abs}(\langle \Pi_m\rangle)^2$.
Therefore, we first determine the connected correlator, by subtraction
of the numerical value of $\mbox{abs}(\langle \Pi_m\rangle)^2$, and then fit
data using Eqs.~(\ref{fit_exp}) or~(\ref{fit_pot}) {\em without}
the additive constant $C$:
\beq
G(r)=A \left( e^{-{\cal M} r}+e^{-{\cal M} (N_\sigma-r)} \right)\;,
\label{fit_exp2}
\eeq
\beq
G(r)=A \left( r^{-\eta}+ (N_\sigma-r)^{-\eta} \right)\;.
\label{fit_pot2}
\eeq

\begin{figure}[htbp]
\centering
\hspace{-1.3cm}
\includegraphics[width=0.75\textwidth,angle=0]{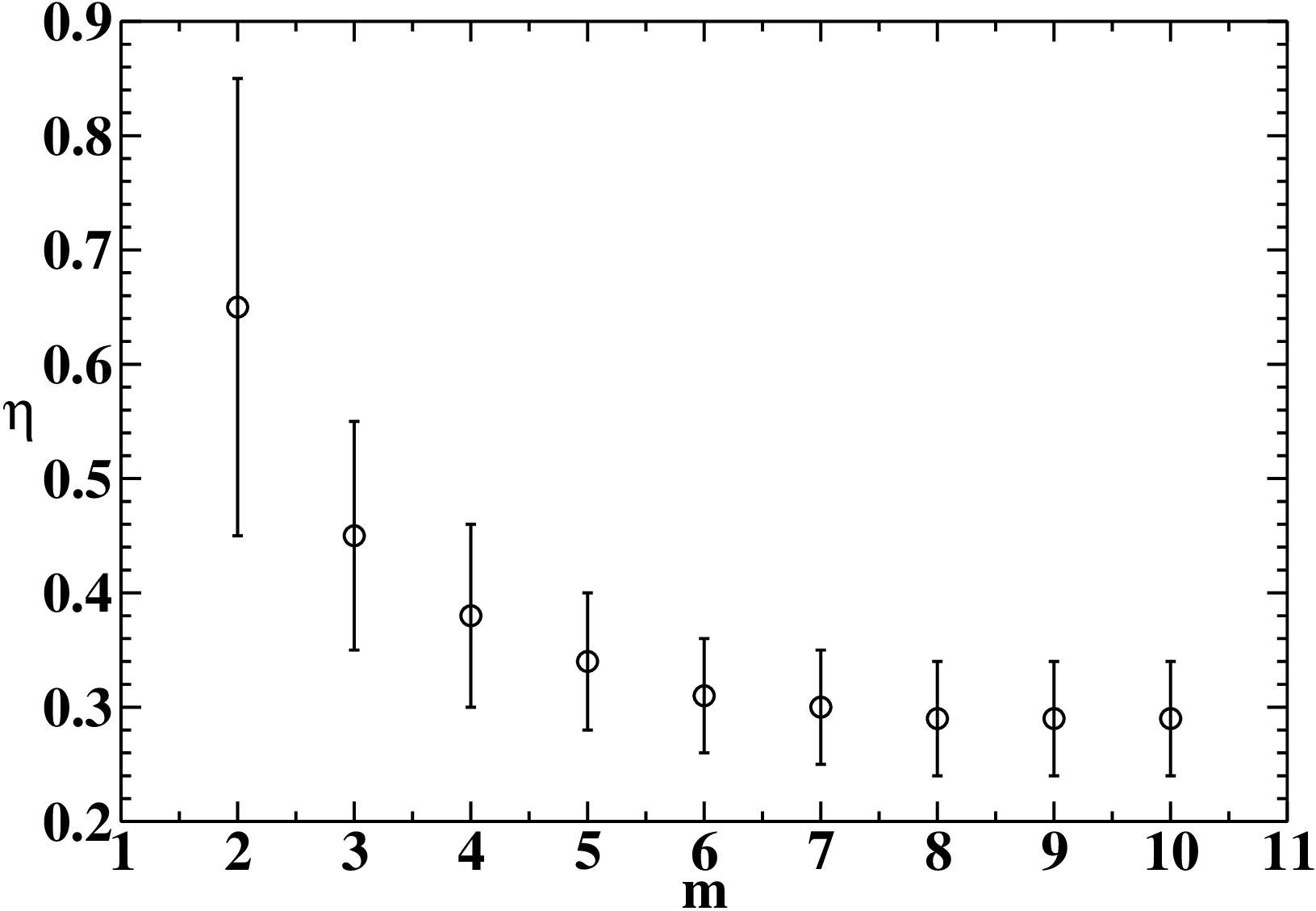}
\caption[]{$\eta$ versus $m$ on a $64^2 \times 8$ lattice at $\beta=0.25$ 
(confined phase); the simulation parameters are as in 
Figure~\ref{G_beta_0.25}.}
\label{eta_vs_m}
\end{figure}

In Figure~\ref{eta_vs_m} we show the value of $\eta$ obtained on the lattice 
with $N_\sigma=64$ by a fit with Eq.~(\ref{fit_pot2}) for different values 
of $m$. Simulations on the lattice with $N_\sigma=32$ do not permit
to obtain reliable estimates, since results are not stable.

\begin{figure}[htbp]
\centering
\hspace{-1.3cm}
\includegraphics[width=0.75\textwidth,angle=0]
{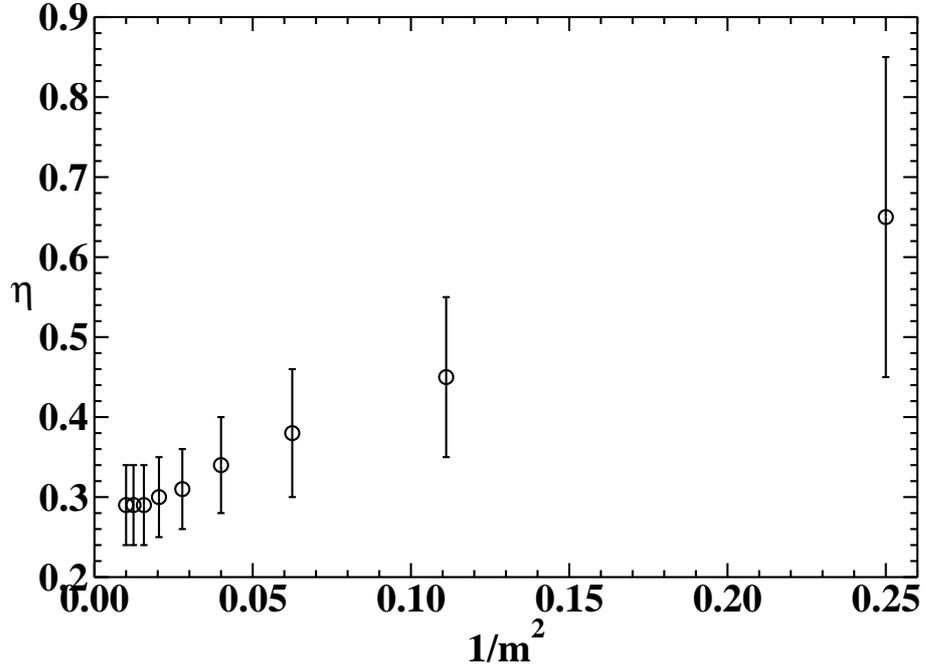}
\caption[]{$\eta$ versus $1/m^2$ on a $64^2 \times 8$ lattice at $\beta=0.25$ 
(confined phase); the simulation parameters are as in 
Figure~\ref{G_beta_0.25}.}
\label{eta_vs_m2}
\end{figure}

It is interesting to note that the behavior of $\eta$ is the one predicted
by Eq.~(\ref{etalattice}), except for an additive constant, since we find 
$\eta = A/m^2+B$ (see Figure~\ref{eta_vs_m2}); for the lattice with 
$N_\sigma=64$, we find $A=1.59(64)$ and $B=0.270(27)$, with $\chi^2$/d.o.f. 
$\sim 0.01$. Note that the tree level theoretical value 
is $\eta=N_{\tau}/({m^2 4 \pi \beta})=2.5464\ldots/m^2$.

If we try to fit data with Eq.~(\ref{fit_exp2}), for both lattices the result 
is a value of $a{\cal M}$ that depends weakly from $m$ and it
is of the order $1/N_\sigma$; this is a strong evidence 
that the system is in a critical region with a ``correlation length'' 
$\xi/a \propto N_\sigma$, that is for $0<T<T_c$ the system is always critical. 
This is a typical feature of the BKT transition.

The error bars on $a{\cal M}$ and $\eta$ are not those resulting from the fit,
which would be underestimated owing to the correlation among the values 
of $G(r)$ at different $r$'s. They are estimated, instead, through the behavior 
of the {\em effective} $\eta$ and $a{\cal M}$, built from suitable ratios 
of $G(r)$ in such a way that the dependence on the parameters $A$ and $B$ 
disappears. This method allowed also to cross-check the results of the fit.

\subsection{Chiral condensate}

\begin{figure}[htbp]
\centering
\hspace{-1.3cm}
\includegraphics[width=0.75\textwidth,angle=0]
{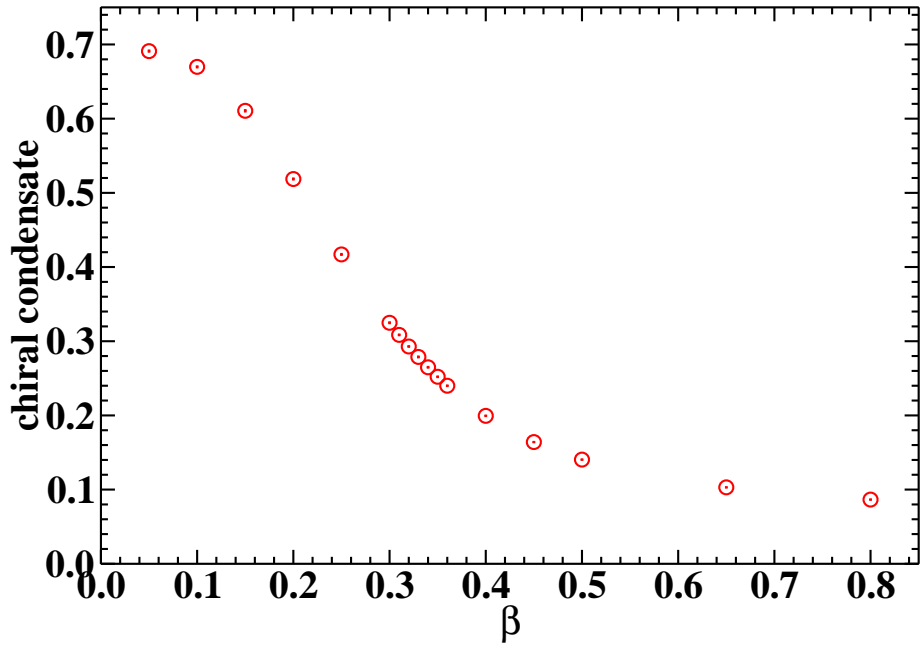}
\caption[]{Chiral condensate versus $\beta$ on a $12^2 \times 8$ lattice; the 
simulation parameters are $aM=0.05$, $\omega=0.1$, $\delta t=0.03$, MCR=50, 
FOM=1000, NMS=10000.}
\label{chiral}
\end{figure}

\begin{figure}[htbp]
\centering
\hspace{-1.3cm}
\includegraphics[width=0.75\textwidth,angle=0]
{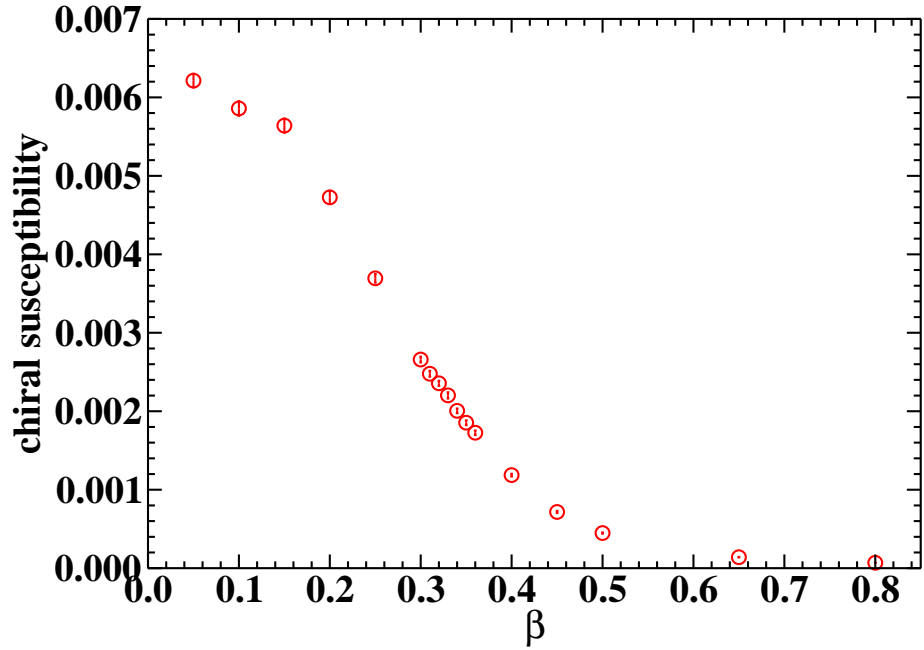}
\caption[]{Chiral susceptibility versus $\beta$ on a $12^2 \times 
8$ lattice; the simulation parameters are as in Figure~\ref{chiral}.}
\label{susc_chiral}
\end{figure}

The results on the chiral condensate are presented in Figures~\ref{chiral} 
and~\ref{susc_chiral}. Data show a smooth transition for increasing $\beta$, 
which supports the argument that there is a chiral transition coinciding
with the confinement-deconfinement one. Results in favor of this conclusion
can be found also in Refs.~\cite{Kogut:1992ny,Aitchison:2000th}.

One would expect, however, that the presence of a transition should be
accompanied by a peak in the susceptibility. This seems to be not the 
case in our analysis -- see Figure~\ref{susc_chiral}. A possible reason for 
this unexpected result could be the small volume used.

\subsection{Monopole density}

\begin{figure}[htbp]
\centering
\hspace{-1.3cm}
\includegraphics[width=0.75\textwidth,angle=0]
{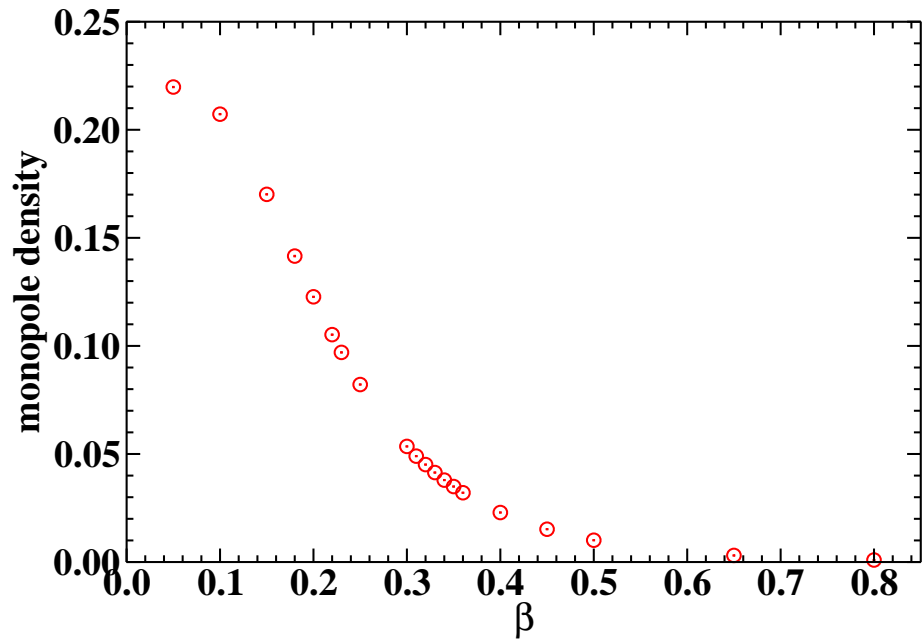}
\caption[]{Monopole density versus $\beta$ on a $12^2 \times 8$ lattice; 
the simulation parameters are $aM=0.05$, $\omega=0.1$, $\delta t=0.03$, MCR=50,
FOM=1000, NMS=10000.}
\label{monopole}
\end{figure}

\begin{figure}[htbp]
\centering
\hspace{-1.3cm}
\includegraphics[width=0.75\textwidth,angle=0]
{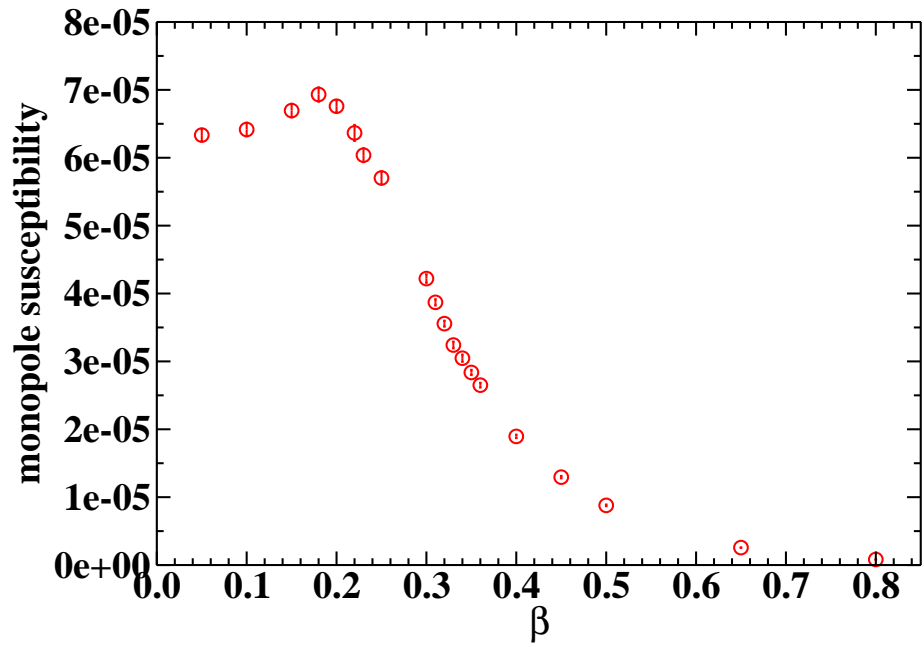}
\caption[]{Monopole density susceptibility versus $\beta$ on a $12^2 \times 8$ 
lattice; the simulation parameters are as in Figure~\ref{monopole}.}
\label{susc_monopole}
\end{figure}

The behavior of the monopole density and of its susceptibility with 
$\beta$ is shown in Figures~\ref{monopole} and~\ref{susc_monopole}, 
respectively. The result is somewhat surprising: the monopole 
density decreases very rapidly in the same region where the Polyakov
loop operators show a fast change, but the peak in the susceptibility is 
located at a different value of the coupling constant, i.e. at 
$\beta_{\rho} \simeq 0.18$.

\section{Conclusions and outlook}
\label{sec:conclusions}

In this paper we have presented an extended numerical analysis on the
confinement-deconfinement transition in non-compact QED$_3$ with massive
fermions at finite temperature. 

We have studied the system for a given value of the fermion mass on lattices
with different extensions. We have found compelling evidence that there 
is indeed a transition temperature from a high-temperature phase where 
fractional charges are deconfined or Debye-screened and a low-temperature
phase with Coulomb-confinement. To detect the transition point we have adopted 
some suitable lattice operators, sensitive to the breaking of the underlying
symmetry of the system, the $\mathbb{Z}$ symmetry. The wall-wall correlation 
of these operators has permitted to characterize the two phases: the 
confined one exhibits power law fall-off with the distance, whereas the 
deconfined one shows exponential decay.

There are several indications that the transition and the phase structure 
are compatible with the suggestions of 
Refs.~\cite{Grignani:1995iv,Grignani:1995hx}:
\begin{itemize}
\item the critical temperature found by numerical simulations is in
remarkable agreement with the one estimated for large fermion mass
in Ref.~\cite{Grignani:1995iv} and, in its turn, consistent with 
the BKT scenario, $T_c=g^2/(8\pi)$; this agreement may be either an 
indication of smooth mass dependence of the effective action used in
Ref.~\cite{Grignani:1995iv} or an accidental fact; in both cases, 
a systematic study of the dependence of our results on the
fermionic mass should be performed; this is, however, beyond the scope of 
the present work;

\item two phases are clearly seen, one where the correlator of the order 
parameter exhibits a fall-off with the distance with a power law, the other 
where the fall-off is exponential; the fact that the power law is valid well 
inside the confined phase and not only on the critical point is an indication 
in favor of the BKT scenario;

\item the scaling of the index $\eta$ for the power law fall-off
with the fractional electric charge is in agreement with the prediction from 
Ref.~\cite{Grignani:1995hx} (except for an additive constant), which supports 
the BKT scenario.

\end{itemize}

There is an indication that the chiral transition has a relation with the
deconfinement transition, which should, however, be confirmed by an analysis
on larger lattices. 

If one defines monopoles on the lattice in the non-compact theory as in
the compact theory, one can see that their density across the deconfinement
transition shows a sharp change. Surprisingly enough, the susceptibility of the
monopole density shows a peak at a smaller temperature than the deconfinement
one.

The future development of this work includes a finite-size scaling analysis 
near the transition temperature, in order to achieve an accurate enough 
determination of the critical indices and to conclude that the transition is 
definitely BKT. Moreover, simulations for different values of the fermion mass 
could allow to explore the phase diagram of the theory in the space of the 
mass, temperature and coupling parameters and to get a unifying view of the 
statistical properties of the system.

\vspace{0.5cm} {\bf \large Acknowledgments}

We thank P.~Sodano for drawing our attention to the subject of this work,
reading the manuscript and suggesting interesting further developments.
We are grateful to D.~Giuliano for reading the manuscript and making
valuable comments.

\newpage

\newpage

\newpage

\newpage

\newpage

\newpage

\newpage

\newpage

\newpage

\newpage

\newpage
\clearpage

\newpage

\newpage

\newpage

\end{document}